\pdfoutput=1

\documentclass[12pt,a4paper]{article}

\usepackage{ifthen} 
\newboolean{pdflatex}
\setboolean{pdflatex}{true} 

\newboolean{articletitles}
\setboolean{articletitles}{true} 

\newboolean{uprightparticles}
\setboolean{uprightparticles}{true} 

\newboolean{inbibliography}
\setboolean{inbibliography}{false} 

\usepackage[top=1in, bottom=1.25in, left=1in, right=1in]{geometry}

\usepackage{microtype}
\usepackage{lineno}  
\usepackage{xspace} 
\usepackage{caption} 

\usepackage{graphicx}  
\usepackage{color}
\usepackage{colortbl}
\graphicspath{{./figs/}} 

\usepackage{amsmath} 
\usepackage{amssymb}
\usepackage{amsfonts}
\usepackage{upgreek} 

\newcommand*\patchAmsMathEnvironmentForLineno[1]{%
\expandafter\let\csname old#1\expandafter\endcsname\csname #1\endcsname
\expandafter\let\csname oldend#1\expandafter\endcsname\csname
end#1\endcsname
 \renewenvironment{#1}%
   {\linenomath\csname old#1\endcsname}%
   {\csname oldend#1\endcsname\endlinenomath}%
}
\newcommand*\patchBothAmsMathEnvironmentsForLineno[1]{%
  \patchAmsMathEnvironmentForLineno{#1}%
  \patchAmsMathEnvironmentForLineno{#1*}%
}
\AtBeginDocument{%
\patchBothAmsMathEnvironmentsForLineno{equation}%
\patchBothAmsMathEnvironmentsForLineno{align}%
\patchBothAmsMathEnvironmentsForLineno{flalign}%
\patchBothAmsMathEnvironmentsForLineno{alignat}%
\patchBothAmsMathEnvironmentsForLineno{gather}%
\patchBothAmsMathEnvironmentsForLineno{multline}%
\patchBothAmsMathEnvironmentsForLineno{eqnarray}%
}

\usepackage{hyperref}    
\usepackage[all]{hypcap} 

\def\lhcb {\mbox{LHCb}\xspace}

\def\MagUp {\mbox{\em Mag\kern -0.05em Up}\xspace}

\ifthenelse{\boolean{uprightparticles}}%
{

 \def\Pmu         {\ensuremath{\upmu}\xspace}

 \def\Ptau        {\ensuremath{\uptau}\xspace}

 \def\PDelta      {\ensuremath{\Delta}\xspace}                 
 \def\PXi      {\ensuremath{\Xi}\xspace}                 
 \def\PLambda      {\ensuremath{\Lambda}\xspace}                 
 \def\PSigma      {\ensuremath{\Sigma}\xspace}                 
 \def\POmega      {\ensuremath{\Omega}\xspace}                 
 \def\PUpsilon      {\ensuremath{\Upsilon}\xspace}                 
 
 \def\PB      {\ensuremath{\mathrm{B}}\xspace}                 
                  
 \def\PD      {\ensuremath{\mathrm{D}}\xspace}

 \def\PK      {\ensuremath{\mathrm{K}}\xspace}

 \def\PZ      {\ensuremath{\mathrm{Z}}\xspace}

 \def\Pe      {\ensuremath{\mathrm{e}}\xspace}

 \def\Pi      {\ensuremath{\mathrm{i}}\xspace}

 \def\Pt      {\ensuremath{\mathrm{t}}\xspace}

}
{

 \def\Pmu         {\ensuremath{\mu}\xspace}

 \def\Ptau        {\ensuremath{\tau}\xspace}

 \mathchardef\PDelta="7101
 \mathchardef\PXi="7104
 \mathchardef\PLambda="7103
 \mathchardef\PSigma="7106
 \mathchardef\POmega="710A
 \mathchardef\PUpsilon="7107
                  
 \def\PB      {\ensuremath{B}\xspace}                 
                  
 \def\PD      {\ensuremath{D}\xspace}

 \def\PK      {\ensuremath{K}\xspace}

 \def\PZ      {\ensuremath{Z}\xspace}

 \def\Pe      {\ensuremath{e}\xspace}

 \def\Pi      {\ensuremath{i}\xspace}

 \def\Pt      {\ensuremath{t}\xspace}

}

\makeatletter
\ifcase \@ptsize \relax
  \newcommand{\miniscule}{\@setfontsize\miniscule{4}{5}}
\or
  \newcommand{\miniscule}{\@setfontsize\miniscule{5}{6}}
\or
  \newcommand{\miniscule}{\@setfontsize\miniscule{5}{6}}
\fi
\makeatother

\DeclareRobustCommand{\optbar}[1]{\shortstack{{\miniscule (\rule[.5ex]{1.25em}{.18mm})}
  \\ [-.7ex] $#1$}}

\def\tquark    {{\ensuremath{\Pt}}\xspace}
\def\tquarkbar {{\ensuremath{\overline \tquark}}\xspace}
\def\ttbar     {{\ensuremath{\tquark\tquarkbar}}\xspace}

  \def\Kbar    {{\kern 0.2em\overline{\kern -0.2em \PK}{}}\xspace}

\def\KorKbar    {\kern 0.18em\optbar{\kern -0.18em K}{}\xspace}

  \def\Dbar    {{\kern 0.2em\overline{\kern -0.2em \PD}{}}\xspace}

\def\DorDbar    {\kern 0.18em\optbar{\kern -0.18em D}{}\xspace}

\def\Bbar    {{\ensuremath{\kern 0.18em\overline{\kern -0.18em \PB}{}}}\xspace}

\def\BorBbar    {\kern 0.18em\optbar{\kern -0.18em B}{}\xspace}

  \def\Y#1S{\ensuremath{\PUpsilon{(#1S)}}\xspace}

\def\Lbar        {{\ensuremath{\kern 0.1em\overline{\kern -0.1em\PLambda}}}\xspace}
\def\LorLbar    {\kern 0.18em\optbar{\kern -0.18em \PLambda}{}\xspace}

\def\to                 {\ensuremath{\rightarrow}\xspace}

\newcommand{\as}{{\ensuremath{\alpha_s}}\xspace}

\def\AT#1     {\ensuremath{A_{\mathrm{T}}^{#1}}\xspace}           

\def\C#1      {\ensuremath{\mathcal{C}_{#1}}\xspace}                       
\def\Cp#1     {\ensuremath{\mathcal{C}_{#1}^{'}}\xspace}                    
\def\Ceff#1   {\ensuremath{\mathcal{C}_{#1}^{\mathrm{(eff)}}}\xspace}        
\def\Cpeff#1  {\ensuremath{\mathcal{C}_{#1}^{'\mathrm{(eff)}}}\xspace}       
\def\Ope#1    {\ensuremath{\mathcal{O}_{#1}}\xspace}                       
\def\Opep#1   {\ensuremath{\mathcal{O}_{#1}^{'}}\xspace}                    

\newcommand{\tev}{\ifthenelse{\boolean{inbibliography}}{\ensuremath{~T\kern -0.05em eV}\xspace}{\ensuremath{\mathrm{\,Te\kern -0.1em V}}}\xspace}
\newcommand{\gev}{\ensuremath{\mathrm{\,Ge\kern -0.1em V}}\xspace}
\newcommand{\mev}{\ensuremath{\mathrm{\,Me\kern -0.1em V}}\xspace}
\newcommand{\kev}{\ensuremath{\mathrm{\,ke\kern -0.1em V}}\xspace}
\newcommand{\ev}{\ensuremath{\mathrm{\,e\kern -0.1em V}}\xspace}
\newcommand{\gevc}{\ensuremath{{\mathrm{\,Ge\kern -0.1em V\!/}c}}\xspace}
\newcommand{\mevc}{\ensuremath{{\mathrm{\,Me\kern -0.1em V\!/}c}}\xspace}
\newcommand{\gevcc}{\ensuremath{{\mathrm{\,Ge\kern -0.1em V\!/}c^2}}\xspace}
\newcommand{\gevgevcccc}{\ensuremath{{\mathrm{\,Ge\kern -0.1em V^2\!/}c^4}}\xspace}
\newcommand{\mevcc}{\ensuremath{{\mathrm{\,Me\kern -0.1em V\!/}c^2}}\xspace}

\def\mum  {\ensuremath{{\,\upmu\rm m}}\xspace}

\def\pb {\ensuremath{\rm \,pb}\xspace}
\def\invpb {\ensuremath{\mbox{\,pb}^{-1}}\xspace}

\def\gsim{{~\raise.15em\hbox{$>$}\kern-.85em
          \lower.35em\hbox{$\sim$}~}\xspace}
\def\lsim{{~\raise.15em\hbox{$<$}\kern-.85em
          \lower.35em\hbox{$\sim$}~}\xspace}

\def\ptot       {\mbox{$p$}\xspace}
\def\pt         {\mbox{$p_{\rm T}$}\xspace}
\def\et         {\mbox{$E_{\rm T}$}\xspace}

\def\evtgen     {\mbox{\textsc{EvtGen}}\xspace}

\def\geant      {\mbox{\textsc{Geant4}}\xspace}

\def\photos     {\mbox{\textsc{Photos}}\xspace}
\def\powheg     {\mbox{\textsc{Powheg}}\xspace}
\def\pythia     {\mbox{\textsc{Pythia}}\xspace}

\def\tell1  {TELL1\xspace}
\def\ukl1   {UKL1\xspace}

\newcommand{\csz}{\ensuremath{\sigma_{\PZ}}\xspace}

\newcommand{\fsr}{\mbox{\textsc{FSR}}\xspace}

\newcommand{\phist}{\ensuremath{\phi^{*}_\eta}\xspace}

\newcommand{\ffsr}{\ensuremath{f_{\textrm{\fsr}}}\xspace}

\newcommand{\herwigpp}{\mbox{\textsc{Herwig++}}\xspace}

\usepackage{cite} 
\usepackage{mciteplus}

\usepackage{longtable} 
\usepackage{multirow}
\usepackage{rotating}
\begin{document}

\renewcommand{\thefootnote}{\fnsymbol{footnote}}
\setcounter{footnote}{1}


\begin{titlepage}
\pagenumbering{roman}

\vspace*{-1.5cm}
\centerline{\large EUROPEAN ORGANIZATION FOR NUCLEAR RESEARCH (CERN)}
\vspace*{1.5cm}
\noindent
\begin{tabular*}{\linewidth}{lc@{\extracolsep{\fill}}r@{\extracolsep{0pt}}}
\ifthenelse{\boolean{pdflatex}}
{\vspace*{-2.9cm}\mbox{\!\!\!\includegraphics[width=.14\textwidth]{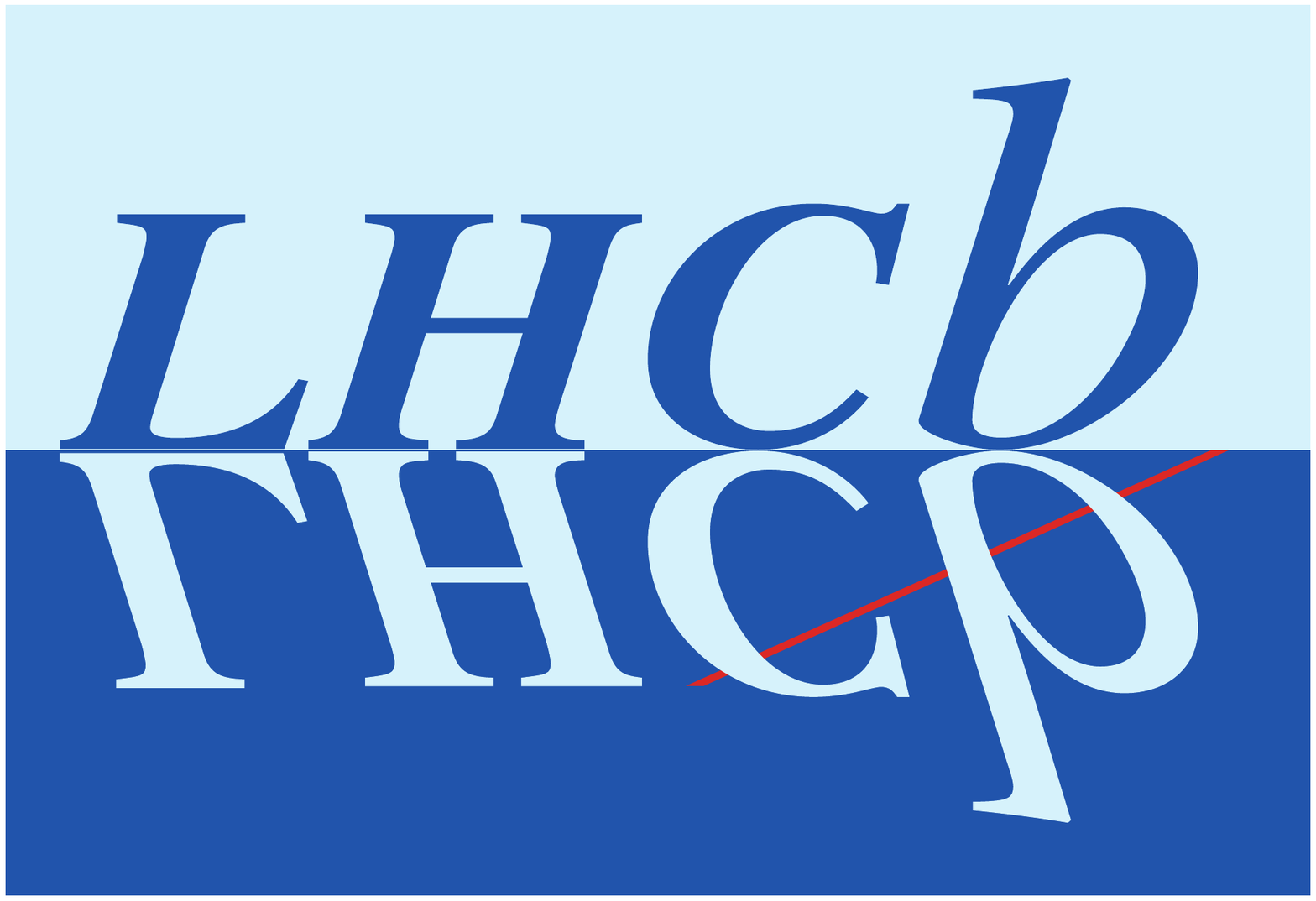}} & &}%
{\vspace*{-1.2cm}\mbox{\!\!\!\includegraphics[width=.12\textwidth]{lhcb-logo.eps}} & &}%
\\
 & & CERN-EP-2016-170 \\  
 & & LHCb-PAPER-2016-021 \\  
 & & September 7, 2016 \\ 
\end{tabular*}

\vspace*{2.0cm}

{\normalfont\bfseries\boldmath\huge
\begin{center}
  Measurement of the forward $\PZ$ boson production cross-section in pp collisions at $\sqrt{s} = 13$~\tev
\end{center}
}

\vspace*{1.0cm}

\begin{center}
The LHCb collaboration\footnote{Authors are listed at the end of this paper.}
\end{center}

\vspace{\fill}

\begin{abstract}
  \noindent
 A measurement of the production cross-section of $\PZ$ bosons in pp collisions at $\sqrt{s} = 13$ TeV is presented using dimuon and dielectron final states in LHCb data. The cross-section is measured for leptons with pseudorapidities in the range $2.0 < \eta < 4.5$, transverse momenta $\pt > 20\gev$ and dilepton invariant mass in the range $60<m(\ell\ell)<120\gev$. The integrated cross-section from averaging the two final states is
\begin{align*}
\csz^{\ell\ell} &= 194.3 \pm 0.9 \pm 3.3 \pm 7.6\pb,
\end{align*}
\noindent where the first uncertainty is statistical, the second is due to systematic effects, and the third is due to the luminosity determination. In addition, differential cross-sections are measured as functions of the \PZ boson rapidity, transverse momentum and the angular variable \phist.
\end{abstract}

\vspace*{0.5cm}

\begin{center}
  Published in JHEP 09 (2016) 136
\end{center}

\vspace{\fill}

{\footnotesize 
\centerline{\copyright~CERN on behalf of the \lhcb collaboration, licence \href{http://creativecommons.org/licenses/by/4.0/}{CC-BY-4.0}.}}
\vspace*{2mm}

\end{titlepage}


\newpage
\setcounter{page}{2}
\mbox{~}
%

%
%
%

\cleardoublepage


\renewcommand{\thefootnote}{\arabic{footnote}}
\setcounter{footnote}{0}



\pagestyle{plain} 
\setcounter{page}{1}
\pagenumbering{arabic}



\section{Introduction}
\label{sec:Introduction}

Measurements are reported of \PZ boson production\footnote{The label \PZ boson is defined to include the effects of virtual photon production and interference terms. The terms electron and muon are also used to refer to both matter and anti-matter species of the particles.} at the LHCb experiment in proton-proton collisions at $\sqrt{s}=13$\tev. The analysis uses a dataset corresponding to an integrated luminosity of $294\pm11$\invpb and considers events where the boson decays either to a dimuon or a dielectron final state. The two final states offer statistically independent samples with largely independent systematic uncertainties. The analysis is performed using similar methods to previous LHCb measurements of electroweak boson production at lower pp collision energies~\cite{LHCb-PAPER-2012-008,LHCb-PAPER-2012-036,LHCb-PAPER-2015-001,LHCb-PAPER-2015-003,LHCb-PAPER-2015-049}. The LHCb detector measures particle production in the forward region; the ATLAS and CMS collaborations have reported similar measurements at $\sqrt{s}=13$\tev~\cite{Aad:2016naf,CMS-PAS-SMP-15-011} in a different kinematic region.

Measurements of electroweak gauge boson production are benchmark tests of Standard Model processes at hadron colliders, and are of interest for constraining the parton distribution functions (PDFs) that describe the structure of the proton. Because of the longitudinal boost required for a \PZ boson to be produced in the forward region, LHCb results are particularly sensitive to effects at low and high values of Bjorken-$x$\cite{Thorne:2008am}, and have been used to constrain global PDF fits~\cite{PDF-MMHT14,PDF-NNPDF30,PDF-CT14}. The $\sqrt{s} = 13$~\tev pp collisions allow LHCb to access lower values of $x$ than previous measurements at 7 and 8~TeV. In addition, the boson transverse momentum (\pt) and $\phi^*_\eta$ distributions can be used to test Monte Carlo modelling of additional higher-order radiation that arises from quantum chromodynamics (QCD). The $\phi^*_\eta$ variable~\cite{Banfi:2010cf} is defined as $\phi^*_\eta \equiv \tan(\phi_\text{acop} / 2) / \cosh(\Delta\eta /2)$, where the acoplanarity angle $\phi_\text{acop}  \equiv \pi - \Delta\phi$ depends on the difference in azimuthal angle of the two leptons, $\Delta\phi$, and $\Delta\eta$ is the difference in pseudorapidity of the two leptons. This variable probes similar physics to that probed by the boson transverse momentum, but with better experimental resolution.

The fiducial region used for the results presented here is the same as in 
previous measurements of \PZ boson production at LHCb~\cite{LHCb-PAPER-2012-008,LHCb-PAPER-2012-029,LHCb-PAPER-2012-036,LHCb-PAPER-2015-001,LHCb-PAPER-2015-003,LHCb-PAPER-2015-049}. Both final-state leptons are required to have $\pt > 20$\gev and pseudorapidity $2.0\nolinebreak<\nolinebreak\eta\nolinebreak<\nolinebreak4.5$.\footnote{This article uses natural units with $c=1$.} The invariant mass of the dilepton pair, $m(\ell\ell)$, is required to be in the range $60\nolinebreak<\nolinebreak m(\ell\ell)\nolinebreak<\nolinebreak120$\gev. The measurements are corrected for final-state radiation to the Born level in quantum electrodynamics (QED), allowing direct comparison of the results in the muon and electron final states, which are reported separately in bins of the boson rapidity, $y_\PZ$, of $\phi^*_\eta$ and, using the dimuon events, as a function of the boson \pt. Cross-sections integrated over the fiducial region (fiducial cross-sections) are also determined using both final states. These are then averaged into a single measurement of the $\PZ\rightarrow \ell\ell$ fiducial cross-section in $\sqrt{s} = 13$~TeV pp collisions.

\section{Detector and simulation}
\label{sec:Detector}

The \lhcb detector~\cite{Alves:2008zz,LHCb-DP-2014-002} is a single-arm forward
spectrometer covering the \mbox{pseudorapidity} range $2<\eta <5$, primarily
designed for the study of particles containing b or c~quarks. The detector includes a high-precision tracking system
consisting of a silicon-strip vertex detector surrounding the pp
interaction region, a large-area silicon-strip detector located
upstream of a dipole magnet with a bending power of about
$4{\mathrm{\,Tm}}$, and three stations of silicon-strip detectors and straw
drift tubes placed downstream of the magnet.
The tracking system provides a measurement of momentum, \ptot, of charged particles with
a relative uncertainty that varies from 0.5\% at low momentum to 1.0\% at 200\gev.
The minimum distance of a track to a primary vertex, the impact parameter, is measured with a resolution of $(15+29/\pt)\mum$,
where the \pt is measured in\,\gev.
Different types of charged hadrons are distinguished using information
from two ring-imaging Cherenkov detectors. 
Photons, electrons and hadrons are identified by a calorimeter system consisting of
scintillating-pad (SPD) and preshower (PS) detectors, an electromagnetic
calorimeter (ECAL) and a hadronic calorimeter (HCAL). Muons are identified by a
system composed of alternating layers of iron and multiwire
proportional chambers.

The online event selection is performed by a trigger, 
which consists of a hardware stage, based on information from the calorimeter and muon
systems, followed by a software stage, which applies a full event
reconstruction. The analysis described here uses triggers designed to select events containing at least one muon or at least one electron. The hardware trigger used for these studies requires that a candidate muon has $\pt > 6$\gev or that a candidate electron has transverse energy $\et > 2.28$\gev. Global event cuts (GEC) are applied in the electron trigger in order to prevent events with high occupancy from dominating the processing time: events only pass the electron trigger if they contain fewer than 450 hits in the SPD detector. No such requirement is made within the muon trigger. The software trigger used here selects events containing a muon candidate with $\pt >12.5$\gev, or an electron candidate with $\pt >15$\gev.

The main challenge with electron reconstruction at LHCb is the energy measurement. The calorimeters at LHCb are optimised for the study of low \et physics, and individual cells saturate for transverse energies greater than approximately 10\gev. Electron reconstruction at LHCb therefore relies on accurate tracking measurements to determine the electron momentum. However, bremsstrahlung photons are often emitted as an electron traverses the LHCb detector, so the measured momentum does not directly correspond to the momentum of the electron produced in the proton-proton collision. These photons are often collinear with the electron and are detected in the same saturated calorimeter cell so that recovery of this emitted photon energy is incomplete. Consequently LHCb accurately determines the direction of electrons, but tends to underestimate their energy by a variable amount, typically around 25\%. Despite these challenges, the excellent angular resolution of electrons provided by the LHCb detector means that measurements using the dielectron final state can be used to complement analyses of angular variables such as rapidity and \phist in the dimuon final state~\cite{LHCb-PAPER-2012-036,LHCb-PAPER-2015-003}.

Simulated pp collisions for the study of reconstruction effects are generated using
\pythia8~\cite{Sjostrand:2007gs,*Sjostrand:2006za}  with a specific \lhcb
configuration~\cite{LHCb-PROC-2010-056}.  Decays of hadronic particles
are described by \evtgen~\cite{Lange:2001uf}, in which final-state
radiation is modelled using \photos~\cite{Golonka:2005pn}. The
interaction of the generated particles with the detector, and its response,
are implemented using the \geant toolkit~\cite{Allison:2006ve, *Agostinelli:2002hh} as described in Ref.~\cite{LHCb-PROC-2011-006}. 

The results reported in this article are compared to fixed-order predictions calculated within perturbative quantum chromodynamics (pQCD) determined using the FEWZ~3.1 generator~\cite{GEN-FEWZ3} at $\mathcal{O}(\alpha_s^2)$, where $\as$ is the coupling strength of the strong force. These predictions do not include electroweak corrections. Predictions are made using MMHT14~\cite{PDF-MMHT14}, NNPDF3.0~\cite{PDF-NNPDF30}, and CT14~\cite{PDF-CT14} PDF sets. In all cases, the factorisation and renormalisation scales are set to the \PZ boson mass. Uncertainties on the fixed-order predictions are evaluated by varying the factorisation and renormalisation scales independently using the seven-point scale variation prescription~\cite{Hamilton:2013fea}, and combining this effect in quadrature with the 68\% CL uncertainties associated with the PDF sets and the value of $\as$. The results are also compared to predictions using the Monash 2013 tune of \pythia8~\cite{Sjostrand:2007gs,*Sjostrand:2006za,Skands:2014pea} and an updated version of the LHCb-specific \pythia8 tune~\cite{LHCb-PROC-2010-056}. In addition, results are compared to predictions from \powheg~\cite{GEN-POWHEG,Alioli:2008gx} at $\mathcal{O}(\as)$ using the NNPDF3.0 PDF set, with the showering implemented using \pythia8. These predictions are calculated using the default \powheg settings and the \pythia8 Monash 2013 tune. The \PZ differential cross-section results are also compared to simulated datasets produced using {\textsc{MadGraph5}}\_aMC@NLO\cite{Alwall:2014hca}. Different schemes are used to match and merge these samples. The MLM~\cite{mlm} sample has leading-order accuracy for the emission of zero, one or two jets; the FxFx~\cite{Frederix:2012ps} sample has next-to-leading-order (NLO) accuracy for zero- or one-jet emission; and the UNLOPS~\cite{Lonnblad:2012ix} sample is accurate at NLO for zero- or one-jet emission and accurate at LO for two-jet emission. Higher jet multiplicities are generated by a parton shower, implemented here using the Monash 2013 tune for \pythia8.

\section{Dataset and event selection}
\label{sec:Selection}
This analysis uses a dataset corresponding to an integrated luminosity of $294\pm11$\invpb recorded by the LHCb experiment in pp collisions at $\sqrt{s} = 13$\tev. This integrated luminosity is determined using the beam-imaging techniques described in Ref.~\cite{LHCb-PAPER-2014-047}. Candidates are selected by requiring two high \pt muons or electrons of opposite charge. Additional requirements are then made to select pure samples; these and the resulting purity are now discussed in turn for the dimuon and dielectron final states.

\subsection{Dimuon final state}
The fiducial requirements outlined in Sect.~\ref{sec:Introduction} are applied as selection criteria for the dimuon final state. In addition, the two tracks are required to satisfy quality criteria and to be identified as muons. At least one of the muons is required to be responsible for the event passing the hardware and software stages of the trigger. The number of selected $\PZ\to\Pmu\Pmu$ candidates is $43\,643$.

Five sources of background are investigated: heavy flavour hadron decays, misidentified hadrons, $\PZ\to\Ptau\Ptau$ decays, \ttbar events, and WW events. Similar techniques to those used in previous analyses are applied to quantify the contribution of each source~\cite{LHCb-PAPER-2015-001,LHCb-PAPER-2015-049}. 
The contribution where at least one muon is produced by the decay of heavy flavour particles is studied by selecting sub-samples where this contribution is enhanced, either by requiring that the muons are not spatially isolated from other activity in the event, or by requiring that the muons are not consistent with a common production point. Studies on these two sub-samples are consistent, and the background contribution is estimated to be $180\pm50$ events.
The contribution from misidentified hadrons is evaluated from the probability with which hadrons are incorrectly identified as muons, and is determined to be $100\pm 13$ events. Following Refs.~\cite{LHCb-PAPER-2012-008,LHCb-PAPER-2015-001,LHCb-PAPER-2015-049}, this evaluation is made with randomly triggered data.
An alternative estimate of the contribution from these sources is found by selecting events where both muons have the same charge, but pass all other selection criteria. The assumption that the charges of the selected muons are uncorrelated for these sources is validated by confirming that the same-sign event yield is compatible with the opposite-sign event yield in background-enriched regions. 
The overall number of same-sign events is 198, with the numbers of $\Pmu^+\Pmu^+$ and $\Pmu^-\Pmu^-$ candidates statistically compatible with each other. 
The difference between this number and the sum of the hadron misidentification and heavy-flavour contributions is assigned as an additional uncertainty on the purity estimate. 
The contribution from $\PZ\to\Ptau\Ptau$ decays where both \Ptau leptons subsequently decay to muons is estimated from \pythia8 simulation to be $30\pm10$ events. The background from muons produced in top-quark decays is determined from simulation normalised using the measurement of the cross-section for top-pair production measured at the ATLAS experiment~\cite{ATLAS-CONF-2015-049}, and is estimated to be $28\pm10$ events. The background from WW decays is also determined from the simulation and found to be negligible. Overall, the purity of the dataset is estimated to be $\rho^{\Pmu\Pmu} = (99.2\pm0.2)\%$, consistent with purity estimates found in previous LHCb measurements at lower centre-of-mass energies~\cite{LHCb-PAPER-2015-001,LHCb-PAPER-2015-049}. As in these previous measurements, no significant variation of the purity is found as a function of the kinematic variables studied, and so the purity is treated as constant. A systematic uncertainty associated with this assumption is discussed in Sect.~\ref{sec:Systematics}.

\subsection{Dielectron final state}
The dielectron final state requires two opposite-sign electron candidates, using the same selection criteria based on calorimeter energy deposits as previous LHCb analyses~\cite{LHCb-PAPER-2012-008,LHCb-PAPER-2015-003}. Electron candidates are required to have $\pt > 20$\gev and $2.0<\eta<4.5$. A loose requirement is made on the dielectron invariant mass, $m(\Pe\Pe) > 40$~\gev, since many events where the dielectron system is produced with an invariant mass above 60~GeV may be reconstructed at lower mass due to bremsstrahlung. Effects arising from the difference between the fiducial acceptance and the selection requirements will be discussed in Sect.~\ref{sec:acc}. At least one of the electrons is required to be responsible for the event passing the hardware and software stages of the LHCb trigger. In total $16\,395$ candidates are selected.

Backgrounds are determined using similar techniques as in previous analyses~\cite{LHCb-PAPER-2012-008,LHCb-PAPER-2015-003}. A sample of same-sign $\Pe^\pm \Pe^\pm$ combinations, otherwise subject to the same selection criteria as the standard dataset, is used to provide a data-based estimate of the largest backgrounds. Hadrons that shower early in the ECAL and fake the signature of an electron are expected to be the dominant background, and should contribute roughly equally to same-sign and opposite-sign pairs. The contribution from heavy-flavour decays is also expected to contribute approximately equally to same-sign and opposite-sign datasets, and is much smaller than the background due to misidentified hadrons. Overall, $1\,255$ candidate same-sign events are selected, with no significant difference observed between the $\Pe^+\Pe^+$ and $\Pe^-\Pe^-$ datasets. In order to ascertain the reliability of this procedure, a hadron-enriched sample is selected by requiring that one of the electron candidates is associated with a significant energy deposit in the HCAL, suggesting that it is likely to be a misidentified hadron. The numbers of same-sign and opposite-sign pairs satisfying these requirements are found to agree within 6.2\%. Consequently a 6.2\% uncertainty is assigned to the estimated yield of background events, which corresponds to a 0.5\% uncertainty on the signal yield. In addition, simulated background datasets of $\PZ\to\Ptau\Ptau$ decays, \ttbar events and WW events are generated~\cite{Sjostrand:2007gs,*Sjostrand:2006za} and studied similarly to the dimuon final state. These all contribute at the level of 0.1\% or less. 
The overall purity of the electron dataset is found to be  $\rho^{\Pe\Pe} = (92.2\pm0.5)\%$.

\section{Cross-section measurement}
\label{sec:Method}
The \PZ boson production cross-section is measured in bins of $y_\PZ$, \phist, and, for the dimuon final state, in bins of the boson \pt. For the dimuon final state the efficiency is obtained from per-event weights that depend on the kinematics of the muons, whereas for the dielectron final state the reconstruction and detection efficiencies are evaluated within each bin of the distribution. These approaches are validated using simulation. 

The cross-section for the dimuon final state in a particular bin $i$ is determined as
\begin{align*}
\csz^{\Pmu\Pmu}(i) =& \frac{1}{L}\, \rho^{\Pmu\Pmu} \ffsr^{\Pmu\Pmu}(i) f_\text{unf}^{\Pmu\Pmu}(i)  \sum\limits_{j=1}^{N_{\PZ}^{\Pmu\Pmu}(i)} \frac{1}{\varepsilon(\Pmu^{+}_j, \Pmu^{-}_j)},
\end{align*}
\noindent where the index $j$ runs over the candidates contributing to the bin, with the total number of candidates in the bin denoted by $N_{\PZ}^{\Pmu\Pmu}(i)$. The total reconstruction and detection efficiency for a given event $j$, $\varepsilon(\Pmu^{+}_j,\Pmu^{-}_j)$, depends on the kinematics of each muon. The correction factors for final-state radiation (FSR) are denoted by $\ffsr^{\Pmu\Pmu}(i)$. Corrections for resolution effects that cause bin-to-bin migrations, where applicable, which do not change the fiducial cross-section, are denoted by $f_\text{unf}^{\Pmu\Pmu}(i)$. Migration of events in and out of the overall LHCb fiducial acceptance is negligible. The purity, introduced earlier, is denoted $\rho^{\Pmu\Pmu}$. The integrated luminosity is denoted by $L$.

For the dielectron final state the cross-section in a particular bin is determined as
\begin{align*}
\csz^{\Pe\Pe}(i) =&\frac{1}{L}\, \rho^{ee}(i) \ffsr^{{\Pe\Pe}}(i) f^{\Pe\Pe}_\text{MZ}(i)   \frac{N_{\PZ}^{\Pe\Pe}(i)}{\varepsilon^{\Pe\Pe}(i)},
\end{align*}
where $N_{\PZ}^{\Pe\Pe}(i)$ denotes the number of candidates in bin $i$. The efficiency associated with reconstructing the dielectron final state in bin $i$ is $\varepsilon^{\Pe\Pe}(i)$ and the purity is $\rho^{\Pe\Pe}$. The correction for FSR from the electrons is denoted $\ffsr^{\Pe\Pe}(i)$, while $f_\text{MZ}^{\Pe\Pe}(i)$ corrects the measurement for migrations in the dielectron invariant mass into and out of the fiducial region. 

For both final states the total cross-section is obtained by summing over $i$. The various correction factors are discussed below.

\subsection{Efficiency determination}
\label{sec:eff}
For the measurement in the dimuon final state, candidates are assigned a weight associated with the probability of reconstructing each muon, and the correction for any inefficiency is applied on an event-by-event basis. Muon reconstruction efficiencies are determined directly from data using the same tag-and-probe techniques as applied in previous LHCb measurements of high-\pt muons~\cite{LHCb-PAPER-2012-008,LHCb-PAPER-2014-033,LHCb-PAPER-2015-001,LHCb-PAPER-2015-049}. Averaged over the muon kinematic distributions, the track reconstruction efficiency is determined to be 95\%, the muon identification efficiency is determined to be 95\% and the single muon trigger efficiency is 80\%. Since either muon can be responsible for the event passing the trigger, the overall efficiency with which candidates pass the trigger is higher, on average 95\%. These efficiencies are determined as a function of the muon pseudorapidity. Efficiency measurements as a function of other variables, such as the muon \pt and the detector occupancy, are studied as a cross-check, with no significant change in the final results.

For the measurement in the dielectron final state, electron reconstruction efficiencies are determined from data and simulation for each bin of the measurement, using the same techniques applied in previous LHCb measurements of $\PZ\rightarrow\Pe\Pe$ production~\cite{LHCb-PAPER-2012-036,LHCb-PAPER-2015-003}. The use of different techniques to determine efficiencies to those applied in the muon channel provides uncorrelated systematic uncertainties between the two measurements. The efficiency for electrons is factorised into similar components to those applied in the dimuon analysis, though one extra effect is considered. The GEC efficiency determines the probability that the dielectron candidates pass the GEC present in the hardware trigger. There is no such requirement in the dimuon trigger. 
The GEC efficiency for dielectron data is determined from the dimuon data, correcting for small differences in the detector response to muons and electrons.
The average GEC efficiency is 79\% and exhibits a weak dependence on rapidity and \phist. The trigger efficiency is determined directly from data using a tag-and-probe method, and is typically 93\%. The efficiency with which both electrons are identified by the calorimetry is typically 78\% and is determined from simulation that has been calibrated with data. This efficiency exhibits a significant dependence on the boson rapidity, since the LHCb calorimeter acceptance only extends as far as $\eta \approx 4.25$. The track reconstruction and kinematic efficiency describes the efficiency with which electrons that are in the fiducial region are reconstructed with $\pt > 20\gev$. It corrects both for failure to reconstruct a track and for incomplete bremsstrahlung recovery incorrectly reconstructing electrons with \pt below the 20~GeV threshold. This is also determined from simulation calibrated to data, and is on average $48\%$. 

\subsection{Resolution effects}
\label{sec:unf}
The excellent angular resolution of the LHCb detector in comparison to the bin widths means that no significant bin-to-bin migrations occur in the \phist or $y_\PZ$ distributions for either the dimuon or dielectron final states. In addition, net migration in and out of the overall LHCb angular acceptance is negligible. However, small migrations in the boson \pt distribution measured using the dimuon final state are expected at low transverse momenta. These effects are typically of similar size to the statistical uncertainty in each bin. This distribution is therefore unfolded to correct for the impact of these migrations using multiplicative correction factors (defined above as $f_\text{unf}^{\Pmu\Pmu}$) determined for each bin from simulation.

\subsection{Final-state radiation corrections}
\label{sec:FSR}
The data are corrected for the effect of FSR from the leptons, allowing comparison of electron and muon final states. The correction in each bin of the measured differential distributions is taken as the average of the values determined using \herwigpp~\cite{GEN-HERWIGPP} and \pythia8~\cite{Sjostrand:2007gs,*Sjostrand:2006za}. The two generators typically agree at the per-mille level; the mean correction is about 2\% for muons and 5\% for electrons, but dependence is seen as functions of the different kinematic variables studied. The strongest variation is seen as a function of the boson \pt, where the correction varies over the distribution by about 10\%. The corrections applied are tabulated in the \hyperref[appendix_results]{appendix}.

\subsection{Acceptance corrections}
\label{sec:acc}
The acceptance correction $f_\text{MZ}^{\text{ee}}$ is applied for electrons to correct for events which pass the selection but are not in the fiducial acceptance in dilepton mass. This correction factor, typically 0.97, is determined from simulation as in previous analyses~\cite{LHCb-PAPER-2012-036,LHCb-PAPER-2015-003} and cross-checked using data.
No correction is applied for muons, where the fiducial acceptance is identical to the kinematic requirement in the acceptance, and where the experimental resolution is sufficient such that net migrations in and out of the acceptance due to experimental resolution are negligible.

\subsection{Measuring fiducial cross-sections}
The fiducial cross-sections are determined by integrating over the $y_{_\PZ}$ distributions. Since no candidates in the bin $4.25<y_{_\PZ}<4.50$ are observed for electrons, a correction for this bin is evaluated using FEWZ~\cite{GEN-FEWZ3}. This correction is found to be 0.7~pb. The fraction of the fiducial cross-section expected in the bin determined using \pythia8 simulation~\cite{Sjostrand:2007gs,*Sjostrand:2006za} is consistent with this estimate to within 0.1~pb. This is assigned as the uncertainty associated with the contribution from this bin to the fiducial cross-section measured in the dielectron final state. Consistent results are obtained when integrating over \phist or \pt.

\section{Systematic uncertainties}
\label{sec:Syst}

The systematic uncertainties associated with the measurement are estimated using the same techniques as in previous analyses~\cite{LHCb-PAPER-2012-008,LHCb-PAPER-2015-001,LHCb-PAPER-2015-003,LHCb-PAPER-2015-049}. The contributions from different sources are combined in quadrature. The uncertainties on the fiducial cross-section measurement are summarised in Table~\ref{tab:systsum}.

For both muons and electrons, the statistical precisions of the efficiencies are assigned as systematic uncertainties. For muons, the accuracy of the tag-and-probe methods used to determine efficiencies is tested in simulation, and efficiencies calculated using the tag-and-probe method are generally found to match simulated efficiencies at the per-mille level, with the largest difference arising from the determination of the track reconstruction efficiency. An uncertainty of 1\% is assigned to this efficiency for each muon. The method of treating each muon independently and applying the efficiencies as a function of the muon pseudorapidity is also studied in simulation, and is found to be accurate to better than 0.6\%. This is also assigned as a systematic uncertainty. For electrons, the accuracy of the method used to determine the trigger efficiency is studied by applying it to the simulated dataset and comparing the resulting efficiencies to those directly determined in the same dataset: no bias is observed, and no additional uncertainty is assigned. For the electron track reconstruction efficiency the relative performance in data and simulation is studied using a tag-and-probe method and an uncertainty of 1.6\% is assigned. The uncertainty associated with potential mismodelling of the electron identification efficiency is determined by comparing between data and simulation the distributions of calorimeter energy deposits used to identify electrons. The impact of any mismodelling is propagated through the measurement, and an uncertainty of 1.3\% is assigned. Apart from the uncertainties arising from the statistical precision of the efficiency evaluation, these uncertainties are treated as fully correlated between bins. Since the efficiencies are determined using different methods for muons and electrons these uncertainties are taken as uncorrelated between the dimuon and dielectron final states.

The uncertainties on the purity estimates described in Sect.~\ref{sec:Selection} introduce uncertainties on the overall cross-sections of $0.2\%$ for muons and $0.5\%$ uncertainty for electrons, treated as correlated between all bins. For the muon analysis, the purity is assumed to be uniform across all bins. To evaluate the uncertainty associated with this assumption, the purity is allowed to vary in each bin, with the change from the nominal result providing an additional uncertainty at the per-mille level for the differential measurement.

The statistical uncertainty on the FSR corrections is treated as a systematic uncertainty on the corrections. This is combined in quadrature with the difference between the corrections derived using the \herwigpp~\cite{GEN-HERWIGPP} and \pythia8~\cite{Sjostrand:2007gs,*Sjostrand:2006za} simulated datasets. The uncertainties on the FSR corrections are taken as uncorrelated between all bins. 

The dimuon analysis is repeated using a momentum scale calibration and detector alignment determined from $\PZ\rightarrow\Pmu\Pmu$ events, in a similar approach to that documented in Ref.~\cite{LHCb-PAPER-2015-039}. The impact on the measured total cross-section and the differential $y_\PZ$ and \phist measurements is negligible. The mean effect in any bin of transverse momentum is typically 1\% and is not statistically significant. However this is assigned as an additional uncertainty on the differential cross-section in each bin of transverse momentum. While the \PZ boson transverse momentum distribution is not measured in the dielectron final state, the momentum scale plays a larger role in the analysis of the dielectron final state due to the significant effect of bremsstrahlung and migrations in electron \pt across the 20~\gev threshold. The impact of the scale around this threshold is evaluated in the same way as in previous $\PZ\rightarrow\Pe\Pe$ analyses at LHCb~\cite{LHCb-PAPER-2012-008,LHCb-PAPER-2015-003}. A fit to the $\min[\pt(\Pe^+),\pt(\Pe^-)]$ spectrum returns a momentum scale correction factor of $1.000\pm0.005$ for simulation. Propagating this uncertainty on the electron momentum scale onto the cross-section measurement yields an uncertainty of about $0.6\%$, which is treated as correlated between all bins.
\label{sec:Systematics}
\begin{table}[!t]
\begin{center}
\caption{Summary of the relative uncertainties on the \PZ boson total cross-section.}
\label{tab:systsum}
\begin{tabular}{lcc}
\textbf{Source} &  \multicolumn{1}{c}{$\Delta\csz^{\Pmu\Pmu}$ [\%]} & \multicolumn{1}{c}{$\Delta\csz^{\Pe\Pe} [\%]$} \\ \hline
Statistical & 0.5 & 0.9\\\hline
Reconstruction efficiencies & 2.4& 2.4\\
Purity & 0.2 & 0.5\\
FSR  & 0.1 & 0.2\\\hline
Total systematic (excl. lumi.) & 2.4 & 2.5\\
Luminosity & {3.9} & 3.9\\
\end{tabular}
\end{center}
\end{table}

The transverse momentum distribution is unfolded to account for potential migration of events between bins arising from the experimental resolution using correction factors in each bin. A systematic uncertainty on this approach is set by considering the Bayesian method~\cite{DAGOSTINI,ROOUNFOLD} with two iterations as an alternative. The difference between the two approaches is at the per-mille level in each bin and is assigned as the uncertainty. As in previous analyses~\cite{LHCb-PAPER-2015-001,LHCb-PAPER-2015-049}, the unfolding is studied using different models of the underlying distribution, and no significant additional variation is observed.

The only uncertainty treated as correlated between the muon and electron final states is the one associated with the luminosity determination. This uncertainty is determined to be 3.9\% following the procedures used in Ref.~\cite{LHCb-PAPER-2014-047}. The uncertainty on the FSR correction may also be correlated, but is sufficiently small for the effects of such correlation to be negligible. The measurement is performed for the nominal centre-of-mass energy of the colliding beams. This energy was determined to an accuracy of 0.65\% for the 4\tev proton beams used in earlier LHC operations~\cite{BEAM}. No studies have yet been published for the 6.5~\tev proton beams used here, but for calculations performed using the FEWZ generator~\cite{GEN-FEWZ3} at NNLO in pQCD, a 0.65\% shift in the beam and collision energy would correspond to a shift in the fiducial cross-section of 0.9\%. This is not assigned as an additional uncertainty. The correlation matrices for the measurements of the differential cross-section as a function of the \PZ boson rapidity are given in the \hyperref[appendix_results]{appendix}.

\section{Results}
\label{sec:Results}

\begin{figure}[!b]
\begin{center}
\includegraphics[width=.75\textwidth]{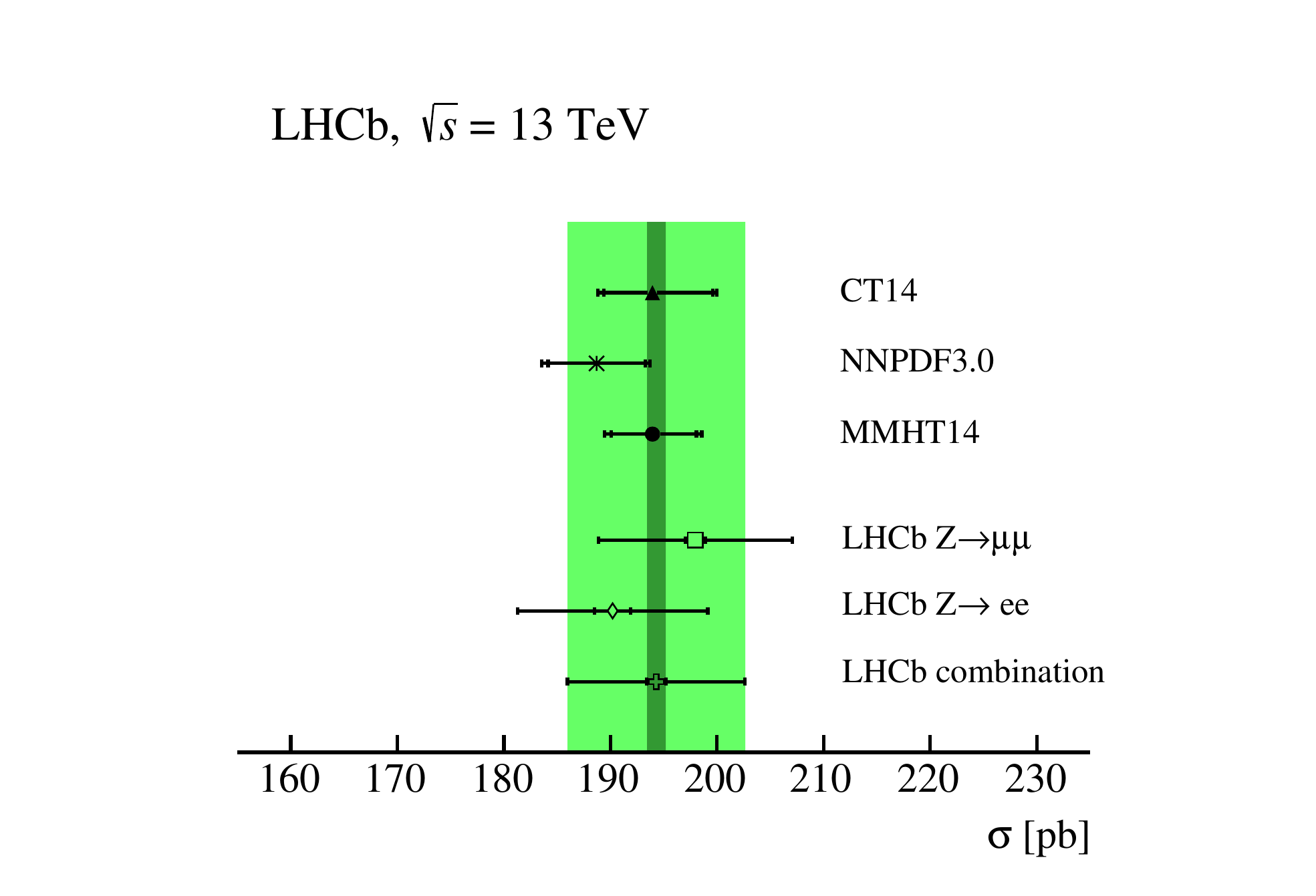}
\caption{The fiducial cross-section compared between theory and data. The bands correspond to the average of the dimuon and dielectron final states, with the inner band corresponding to the statistical uncertainty and the outer band corresponding to the total uncertainty. The top three points correspond to $\mathcal{O}(\alpha_{s}^2)$ predictions with different PDF sets. The inner error bars on these points are due to the PDF uncertainty, with the outer error bars giving the contribution of all uncertainties. The bottom points correspond to the LHCb measurements in the dielectron and dimuon final states and their average, with the inner error bar showing the statistical uncertainty and the outer error bar the total uncertainty.}
\label{fig:totZ}  
\end{center}
\end{figure}

\begin{figure}[!t]
\begin{center}
\includegraphics[width=.75\textwidth]{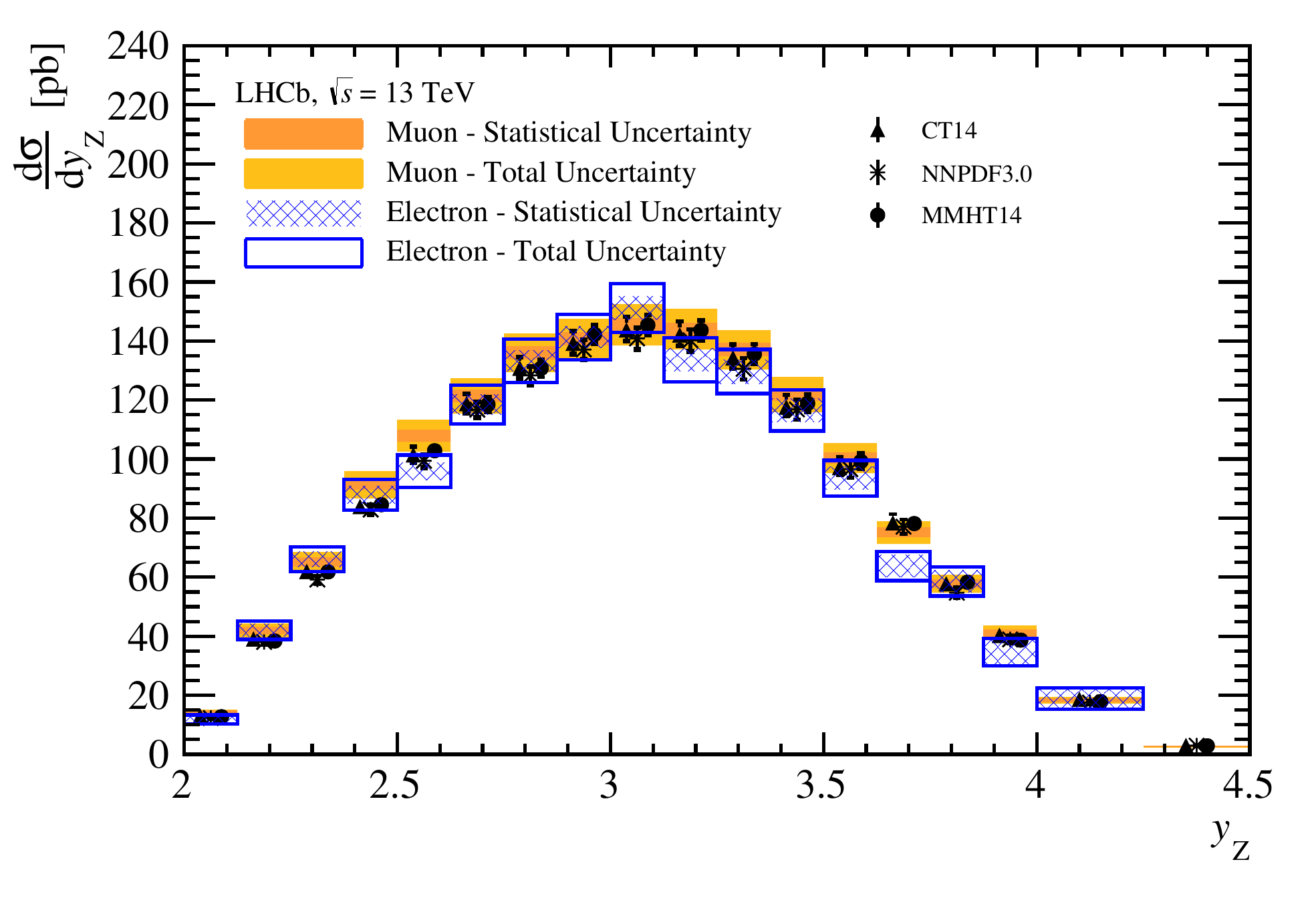}
\includegraphics[width=.75\textwidth]{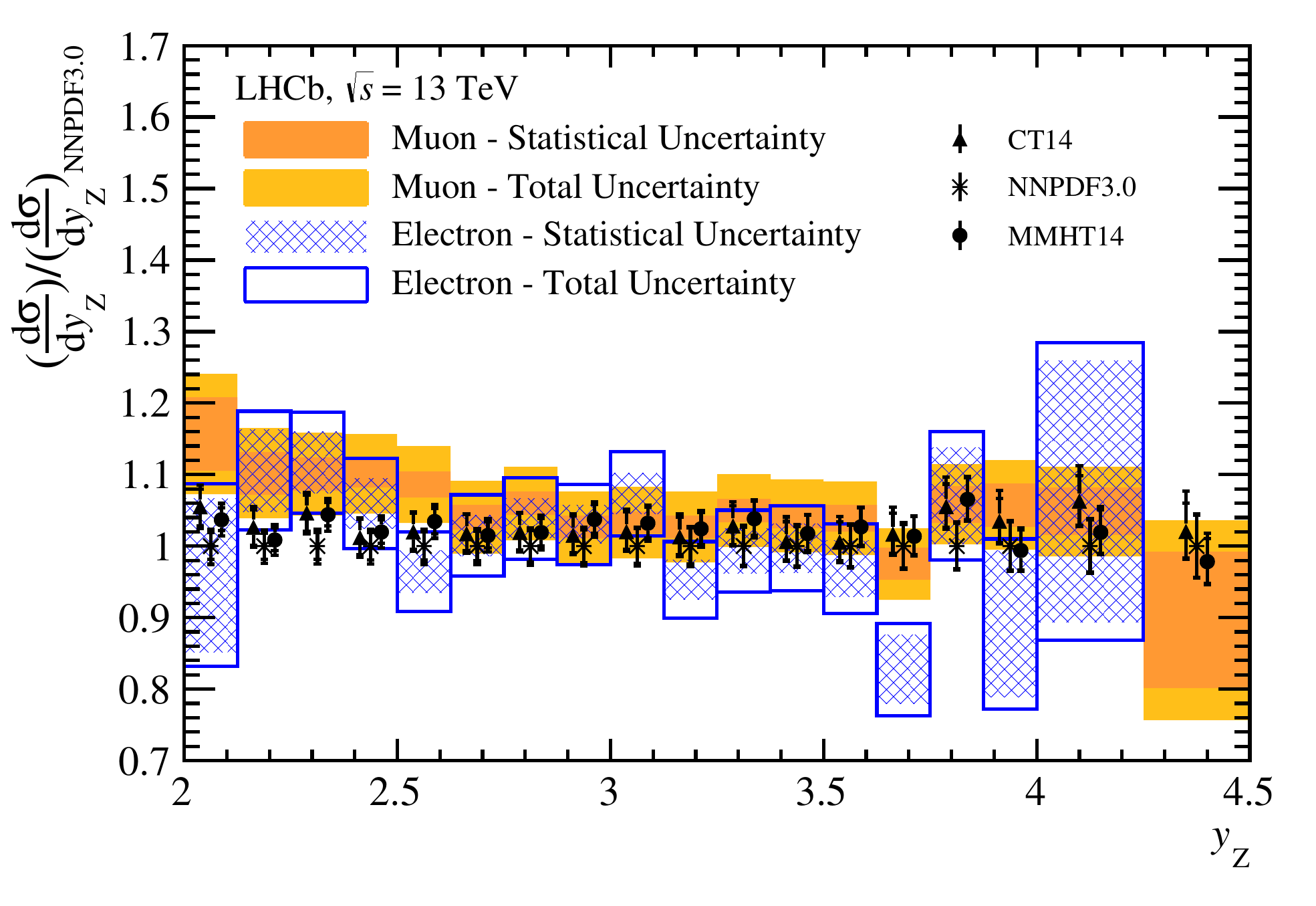}
\caption{The differential cross-section as a function of the \PZ boson rapidity, compared between theory and data. The bands correspond to the data, with the inner band corresponding to the statistical uncertainty and the outer band corresponding to the total uncertainty. The points correspond to $\mathcal{O}(\alpha_{s}^2)$ predictions with different PDF sets. The inner error bars on these points are due to the PDF uncertainty, with the outer error bars giving the contribution of all uncertainties. The different predictions are displaced horizontally within bins to enable ease of comparison. The upper plot shows the differential cross-section, and the lower plot shows the same information as ratios to the central values of the NNPDF3.0 predictions.}
\label{fig:rapZ}  
\end{center}
\end{figure}

\begin{figure}[!t]
\begin{center}
\includegraphics[width=.75\textwidth]{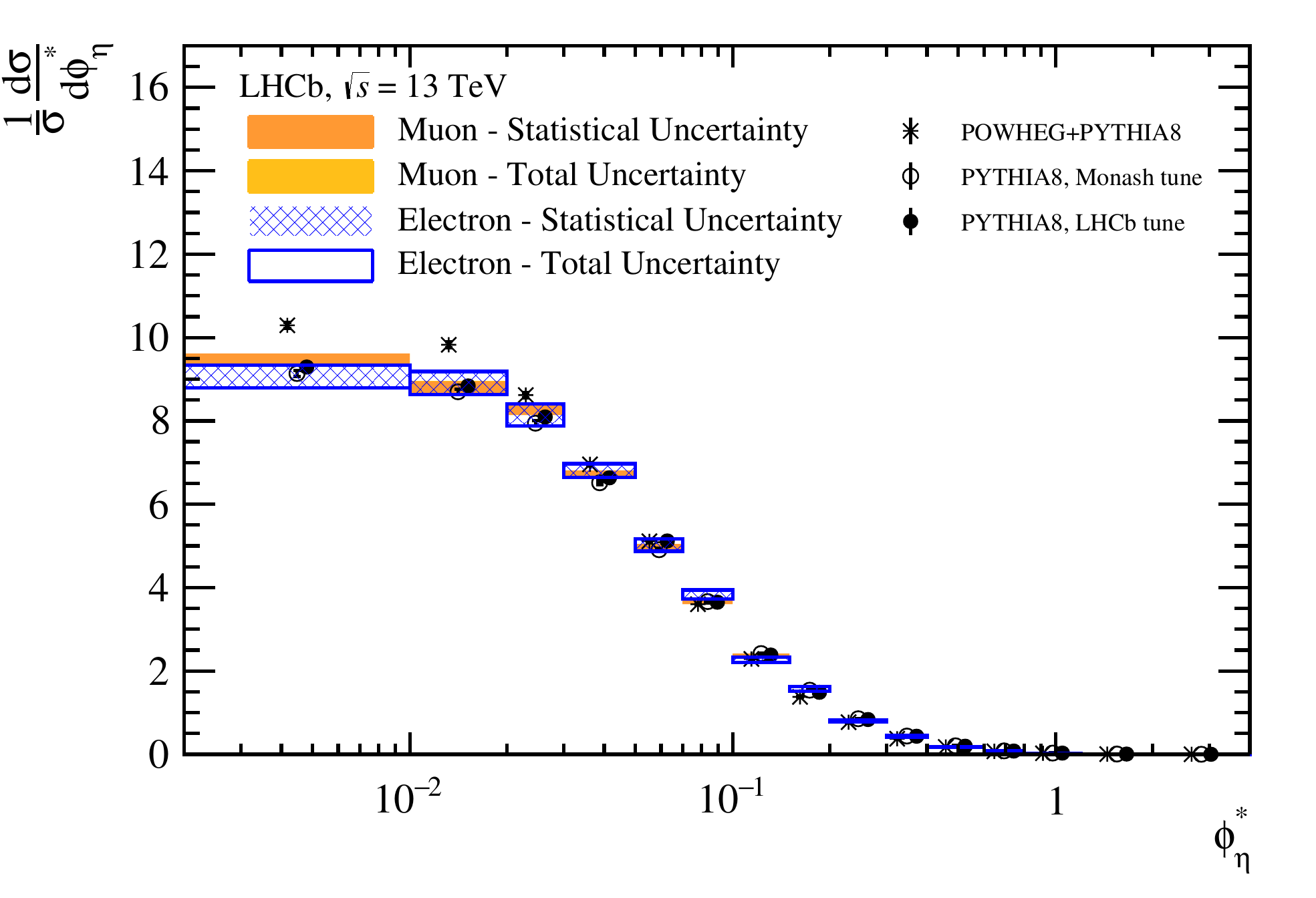}
\includegraphics[width=.75\textwidth]{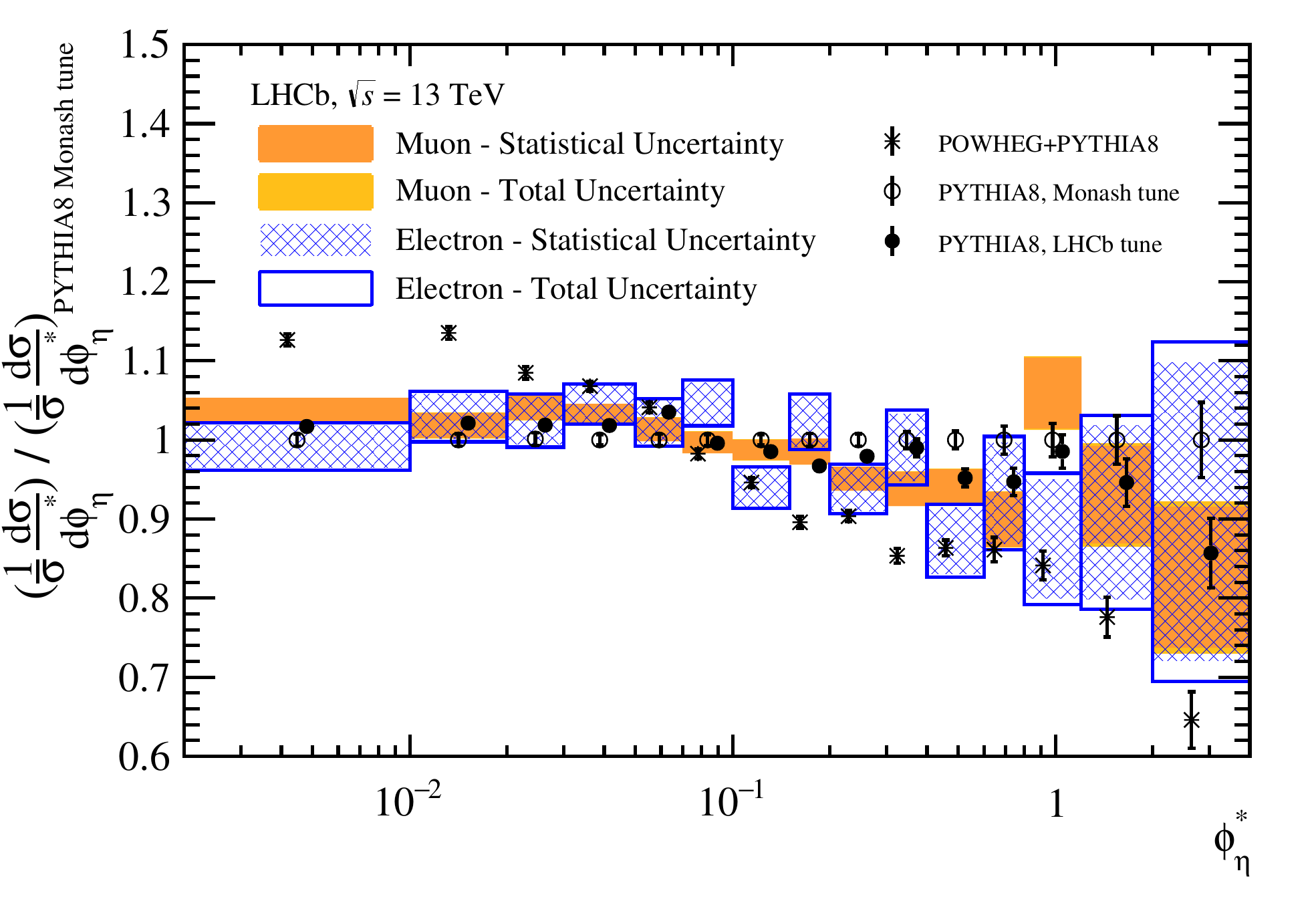}
\caption{The normalised differential cross-section as a function of the \PZ boson $\phi^{*}_\eta$, compared between theory and data. The bands correspond to the data, with the inner band corresponding to the statistical uncertainty and the outer band corresponding to the total uncertainty. The points correspond to the theoretical predictions from the different generators and tunes. The different predictions are displaced horizontally within bins to enable ease of comparison. The upper plot shows the normalised differential cross-section, and the lower plot shows the same information as ratios to the central values of the predictions produced using the Monash 2013 tune of \pythia8. The uncertainties on the theoretical predictions, visible at high \phist, are statistical.}
\label{fig:psZ}  
\end{center}
\end{figure}

\begin{figure}[!t]
\begin{center}
\includegraphics[width=.75\textwidth]{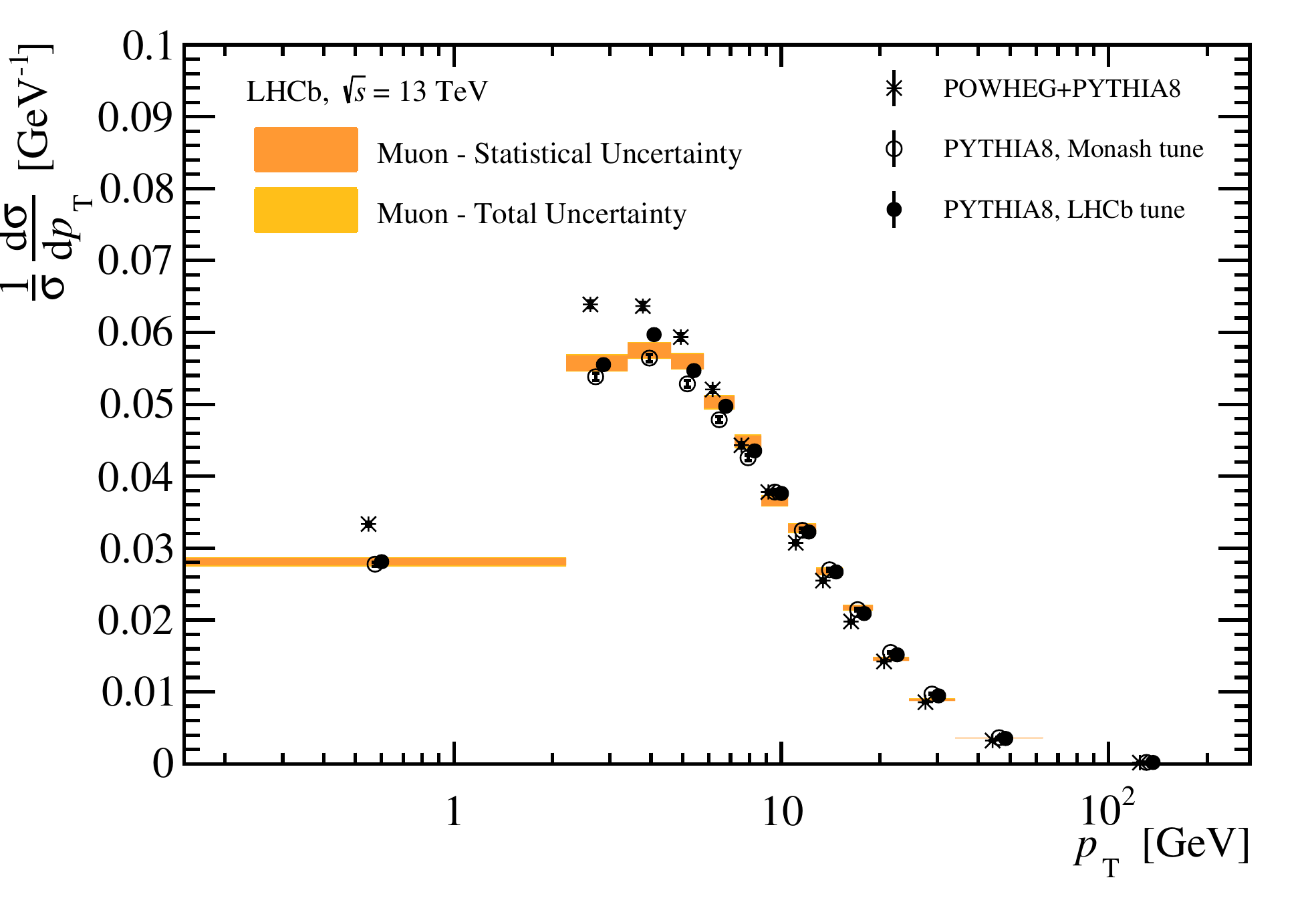}
\includegraphics[width=.75\textwidth]{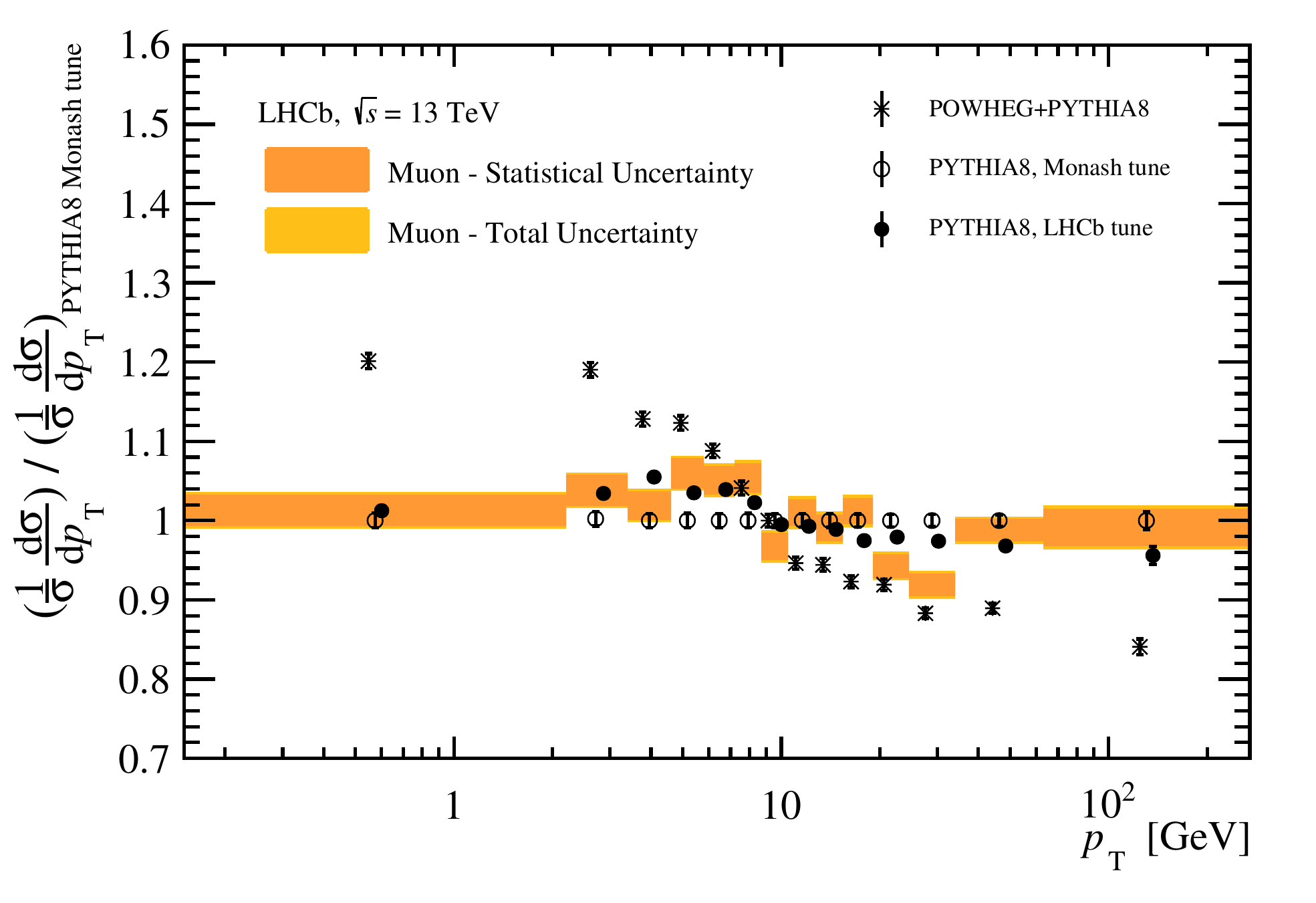}
\caption{The normalised differential cross-section as a function of the \PZ boson transverse momentum, compared between theory and data. The bands correspond to the data, with the inner band corresponding to the statistical uncertainty and the outer band corresponding to the total uncertainty. The points correspond to the theoretical predictions from the different generators and tunes. The different predictions are displaced horizontally within bins to enable ease of comparison. The upper plot shows the normalised differential cross-section, and the lower plot shows the same information as ratios to the central values of the predictions produced using the Monash 2013 tune of \pythia8.}
\label{fig:ptZ}  
\end{center}
\end{figure}

\begin{figure}[!t]
\begin{center}
\includegraphics[width=.75\textwidth]{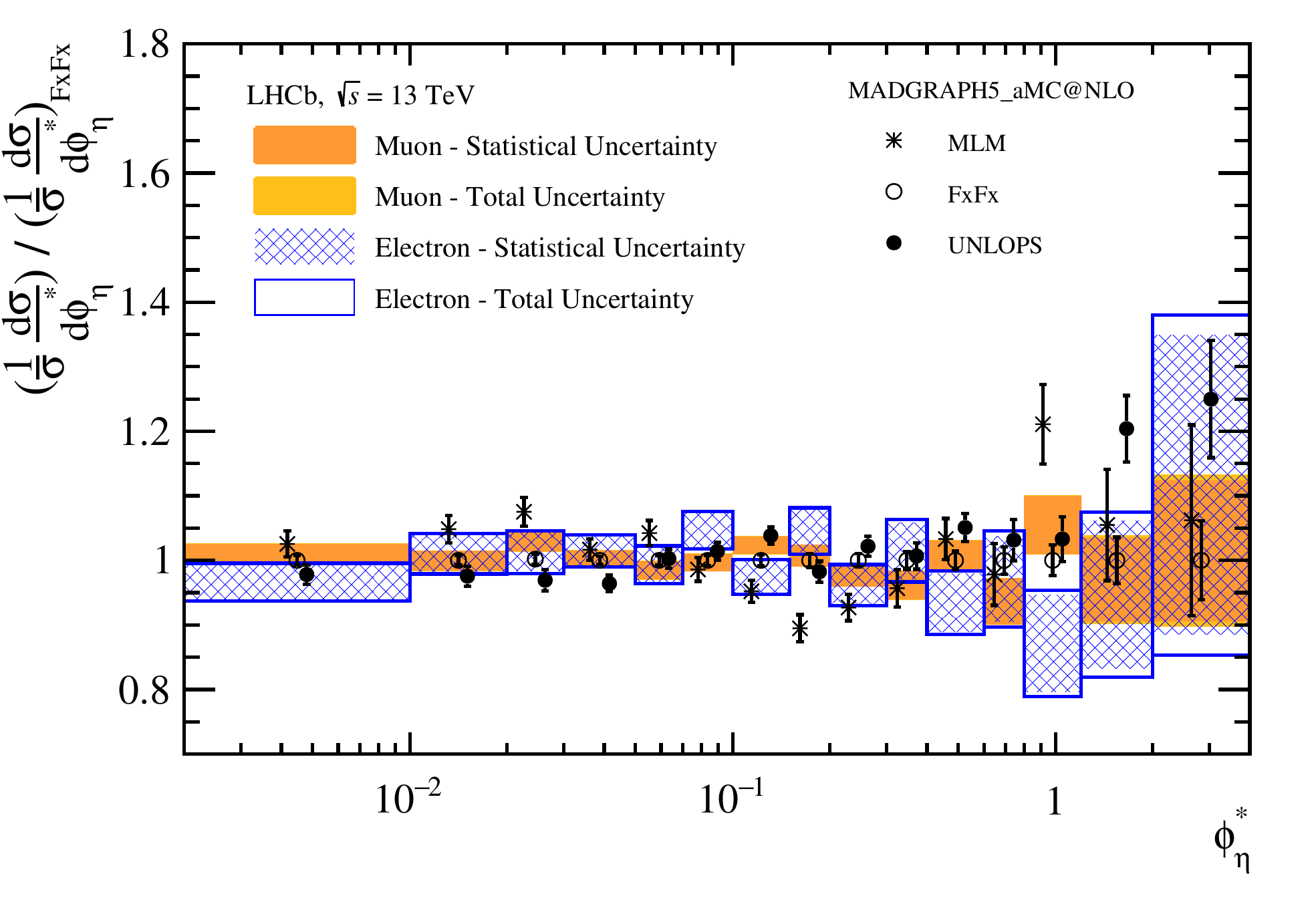}
\includegraphics[width=.75\textwidth]{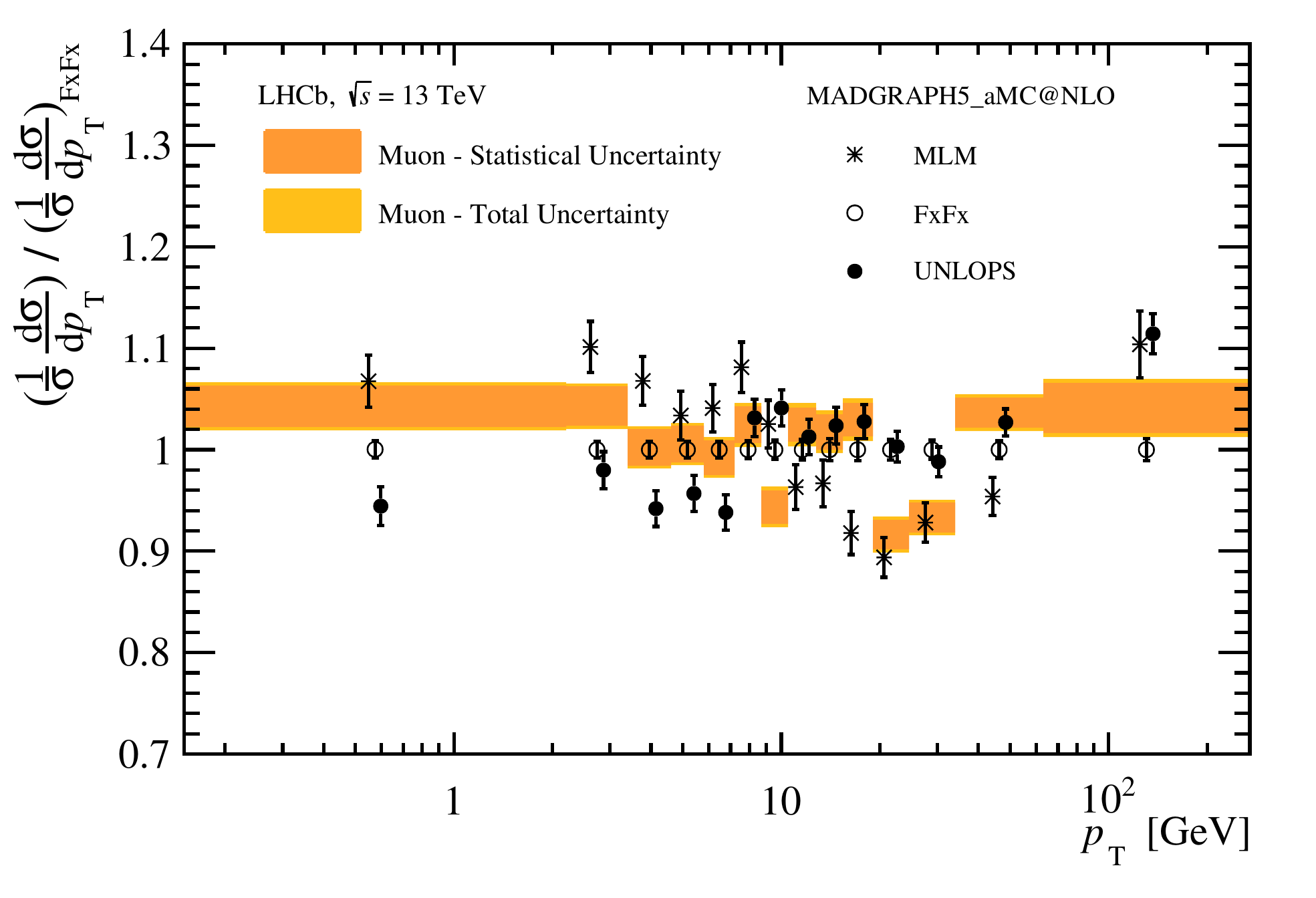}
\caption{The ratio of the normalised differential cross-sections to the predictions evaluated using the FxFx scheme. The bands correspond to the data, with the inner band corresponding to the statistical uncertainty and the outer band corresponding to the total uncertainty. The different predictions are displaced horizontally within bins to enable ease of comparison. Alternative schemes give different predictions, shown as points.All predictions are generated using {\textsc{MadGraph5}}\_aMC@NLO. The uncertainties on the theoretical predictions are statistical. The upper plot shows the $\phist$ distribution, and the lower plot shows the \pt distribution.}
\label{fig:mm}  
\end{center}
\end{figure}

The inclusive \PZ boson cross-section for decays to a dilepton final state with the dilepton invariant mass in the range $60 < m(\ell\ell) < 120$~\gev, and where the leptons have $p_\text{T}\nolinebreak>\nolinebreak20$\gev and $2.0<\eta<4.5$, is measured in $\sqrt{s} = 13\tev$ pp collisions to be 
\begin{align*}
\csz^{\Pmu\Pmu} &= 198.0 \pm 0.9 \pm 4.7 \pm 7.7\pb,\\
\csz^{\Pe\Pe} &= 190.2 \pm 1.7 \pm 4.7 \pm 7.4\pb.
\end{align*}
\noindent The first uncertainties quoted are statistical, the second arise from systematic effects, and the third are due to the accuracy of the luminosity determination. This cross-section is determined at the Born level in QED. Taking the luminosity uncertainty to be fully correlated, the two measurements are consistent at the level of $1.1\,\sigma$, and are linearly combined to give 
\begin{align*}
\csz^{\ell\ell} &= 194.3 \pm 0.9 \pm 3.3 \pm 7.6\pb,
\end{align*}
where the combination minimises the sum of the statistical and systematic uncertaintes in quadrature. The integrated cross-section in the fiducial acceptance and the differential measurement as a function of the \PZ boson rapidity are compared in Figs.~\ref{fig:totZ} and~\ref{fig:rapZ} to the fixed-order predictions for both dimuon and dielectron final states. The measured differential cross-sections are tabulated in the \hyperref[appendix_results]{appendix}. Fixed-order predictions describe the LHCb data well for a range of PDF sets. The measured differential cross-section is slightly larger than the next-to-next-to-leading order pQCD predictions at lower rapidities, in line with observations in Ref.~\cite{CMS-PAS-SMP-15-011}. The differences between the PDF sets, and the PDF uncertainties, are larger than those at lower values of $\sqrt{s}$. Larger LHCb datasets with the uncertainty on the luminosity determination reduced to the level of previous studies (1.2\%) should significantly constrain the PDFs. The differential cross-sections as a function of \pt and \phist, normalised to the total cross-section, are shown in Figs.~\ref{fig:psZ},~\ref{fig:ptZ} and~\ref{fig:mm}. Since the largest systematic effects are independent of these variables, systematic uncertainties largely cancel when these distributions are normalised, and the uncertainties on the normalised distributions are dominated by the statistical components. The LHCb data agree better with \pythia8 predictions than with \powheg+~\pythia\nolinebreak8 predictions, as seen also in previous analyses~\cite{LHCb-PAPER-2015-001,LHCb-PAPER-2012-036}. The LHCb specific tune of \pythia8 does not describe the data significantly better than the Monash 2013 tune. In addition, the data do not favour a particular matching and merging scheme generated using {\textsc{MadGraph5}}\_aMC@NLO.

\section{Conclusions}
\label{sec:Conclusions}
The \PZ production cross-section measured in pp collisions at $\sqrt{s} = 13$\tev is presented using LHCb events where the $\PZ$ boson decays to two muons or two electrons. The cross-section is measured in a fiducial acceptance defined by lepton pseudorapidity in the range $2.0 < \eta < 4.5$, transverse momentum $\pt > 20\gev$, and dilepton invariant mass in the range $60<m(\ell\ell)<120\gev$. The cross-section is measured to be 
\begin{align*}
\csz^{\ell\ell} &= 194.3 \pm 0.9 \pm 3.3 \pm 7.6\pb,
\end{align*}
\noindent where the uncertainties are due to the size of the dataset, systematic effects, and the luminosity determination respectively.
 In addition, the measurement is performed in bins of the \PZ boson rapidity, transverse momentum and \phist. The measurement is compared to theoretical predictions calculated at $\mathcal{O}(\alpha_{s}^2)$ in pQCD as a function of the boson rapidity. The results do not favour any specific parton distribution function, but the differences between the PDF sets suggest that, with more data and a reduction in the uncertainty associated with the luminosity determination, LHCb results will significantly constrain the PDFs. The \phist and boson transverse momentum distributions are compared to theoretical predictions that model higher orders in pQCD in different ways. No significant deviations are seen between the data and the Standard Model.  

\pagebreak

\section*{Acknowledgements}
\noindent We express our gratitude to our colleagues in the CERN
accelerator departments for the excellent performance of the LHC. We
thank the technical and administrative staff at the LHCb
institutes. We acknowledge support from CERN and from the national
agencies: CAPES, CNPq, FAPERJ and FINEP (Brazil); NSFC (China);
CNRS/IN2P3 (France); BMBF, DFG and MPG (Germany); INFN (Italy); 
FOM and NWO (The Netherlands); MNiSW and NCN (Poland); MEN/IFA (Romania); 
MinES and FASO (Russia); MinECo (Spain); SNSF and SER (Switzerland); 
NASU (Ukraine); STFC (United Kingdom); NSF (USA).
We acknowledge the computing resources that are provided by CERN, IN2P3 (France), KIT and DESY (Germany), INFN (Italy), SURF (The Netherlands), PIC (Spain), GridPP (United Kingdom), RRCKI and Yandex LLC (Russia), CSCS (Switzerland), IFIN-HH (Romania), CBPF (Brazil), PL-GRID (Poland) and OSC (USA). We are indebted to the communities behind the multiple open 
source software packages on which we depend.
Individual groups or members have received support from AvH Foundation (Germany),
EPLANET, Marie Sk\l{}odowska-Curie Actions and ERC (European Union), 
Conseil G\'{e}n\'{e}ral de Haute-Savoie, Labex ENIGMASS and OCEVU, 
R\'{e}gion Auvergne (France), RFBR and Yandex LLC (Russia), GVA, XuntaGal and GENCAT (Spain), Herchel Smith Fund, The Royal Society, Royal Commission for the Exhibition of 1851 and the Leverhulme Trust (United Kingdom).

\clearpage

\clearpage

{\noindent\normalfont\bfseries\Large Appendix}

\appendix

\section*{Tabulated results and correlation matrices}
\label{appendix_results}
The FSR corrections used in this analysis are given in Tables~\ref{tab:fsr1},~\ref{tab:fsr3}, and~\ref{tab:fsr2}. The bins are indexed in increasing rapidity, \phist and transverse momentum, and the same binning schemes as in Refs.~\cite{LHCb-PAPER-2015-001,LHCb-PAPER-2015-003,LHCb-PAPER-2015-049} are used. The bin index scheme defined in Tables~\ref{tab:fsr1},~\ref{tab:fsr3}, and~\ref{tab:fsr2} is used throughout the appendix. The differential cross-section results are tabulated in Tables~\ref{tab:rapidity},~\ref{tab:phist} and~\ref{tab:pt}. The correlation matrices are given in Tables~\ref{tab:corr_rap},~\ref{tab:corr_rap_electron},~\ref{tab:corr_ps_mu},~\ref{tab:corr_ps_e}, and~\ref{tab:corr_pt}.

\begin{table}[!h]
\begin{center}
\caption{The FSR correction applied as a function of the boson rapidity.\label{tab:fsr1}}
\begin{tabular}{cc|cc}
Bin index & Bin range& $f_\text{FSR}^{\Pmu\Pmu}$ & $f_\text{FSR}^{\Pe\Pe}$\\\hline
1&2.000-2.125&1.016$\pm$0.005&1.034$\pm$0.003\\
2&2.125-2.250&1.017$\pm$0.004&1.037$\pm$0.005\\
3&2.250-2.375&1.021$\pm$0.002&1.040$\pm$0.002\\
4&2.375-2.500&1.018$\pm$0.002&1.041$\pm$0.002\\
5&2.500-2.625&1.023$\pm$0.003&1.043$\pm$0.002\\
6&2.625-2.750&1.022$\pm$0.003&1.044$\pm$0.004\\
7&2.750-2.875&1.022$\pm$0.002&1.047$\pm$0.004\\
8&2.875-3.000&1.023$\pm$0.003&1.048$\pm$0.002\\
9&3.000-3.125&1.026$\pm$0.002&1.051$\pm$0.002\\
10&3.125-3.250&1.026$\pm$0.002&1.051$\pm$0.002\\
11&3.250-3.375&1.025$\pm$0.004&1.055$\pm$0.001\\
12&3.375-3.500&1.026$\pm$0.005&1.053$\pm$0.003\\
13&3.500-3.625&1.027$\pm$0.002&1.049$\pm$0.005\\
14&3.625-3.750&1.024$\pm$0.002&1.051$\pm$0.007\\
15&3.750-3.875&1.021$\pm$0.003&1.045$\pm$0.004\\
16&3.875-4.000&1.019$\pm$0.019&1.038$\pm$0.011\\
17&4.000-4.250&1.034$\pm$0.014&1.061$\pm$0.013\\
18&4.250-4.500&1.046$\pm$0.119&
\end{tabular}
\end{center}
\end{table}

\begin{table}[!t]
\begin{center}
\caption{The FSR correction applied as a function of \phist.\label{tab:fsr3}}
\begin{tabular}{cc|cc}
Bin index &Bin range& $f_\text{FSR}^{\Pmu\Pmu}$ & $f_\text{FSR}^{\Pe\Pe}$\\\hline
1&0.00-0.01&1.034$\pm$0.002&1.057$\pm$0.002\\
2&0.01-0.02&1.035$\pm$0.002&1.057$\pm$0.001\\
3&0.02-0.03&1.028$\pm$0.001&1.054$\pm$0.001\\
4&0.03-0.05&1.027$\pm$0.002&1.050$\pm$0.002\\
5&0.05-0.07&1.022$\pm$0.002&1.048$\pm$0.001\\
6&0.07-0.10&1.018$\pm$0.003&1.041$\pm$0.002\\
7&0.10-0.15&1.015$\pm$0.004&1.040$\pm$0.004\\
8&0.15-0.20&1.016$\pm$0.001&1.038$\pm$0.003\\
9&0.20-0.30&1.012$\pm$0.003&1.039$\pm$0.002\\
10&0.30-0.40&1.014$\pm$0.003&1.042$\pm$0.003\\
11&0.40-0.60&1.017$\pm$0.005&1.042$\pm$0.002\\
12&0.60-0.80&1.021$\pm$0.004&1.044$\pm$0.007\\
13&0.80-1.20&1.027$\pm$0.010&1.044$\pm$0.004\\
14&1.20-2.00&1.028$\pm$0.008&1.048$\pm$0.007\\
15&2.00-4.00&1.002$\pm$0.041&1.080$\pm$0.023

\end{tabular}
\end{center}
\end{table}

\begin{table}[!t]
\begin{center}
\caption{The FSR correction applied as a function of the boson transverse momentum.\label{tab:fsr2}}
\begin{tabular}{cc|c}
Bin index & Bin range [GeV] &$f_\text{FSR}^{\Pmu\Pmu}$\\\hline
1&0.0-2.2&1.090$\pm$0.004\\
2&2.2-3.4&1.075$\pm$0.002\\
3&3.4-4.6&1.062$\pm$0.003\\
4&4.6-5.8&1.045$\pm$0.003\\
5&5.8-7.2&1.029$\pm$0.001\\
6&7.2-8.7&1.014$\pm$0.005\\
7&8.7-10.5&1.002$\pm$0.007\\
8&10.5-12.8&0.990$\pm$0.008\\
9&12.8-15.4&0.984$\pm$0.005\\
10&15.4-19.0&0.976$\pm$0.008\\
11&19.0-24.5&0.980$\pm$0.005\\
12&24.5-34.0&1.007$\pm$0.002\\
13&34.0-63.0&1.035$\pm$0.001\\
14&63.0-270.0&1.064$\pm$0.004
\end{tabular}
\end{center}
\end{table}

\begin{sidewaystable}
\begin{center}
\caption{The measured differential cross-sections as a function of the boson rapidity. The first uncertainty is due to the size of the dataset, the second is due to experimental systematic uncertainties, and the third is due to the luminosity.}
\label{tab:rapidity}
\begin{tabular}{c|ccccccc|ccccccc}
Bin index & \multicolumn{7}{c|}{$\rm{d}\csz^{\Pmu\Pmu}/\rm{d}y_\PZ$  [pb]} & \multicolumn{7}{c}{$\rm{d}\csz^{\rm{ee}}/\rm{d}y_\PZ$  [pb]}\\\hline
1&14.2&$\pm$&0.7&$\pm$&0.5&$\pm$&0.6&11.8&$\pm$&1.3&$\pm$&0.7&$\pm$&0.5\\
2&41.9&$\pm$&1.2&$\pm$&1.2&$\pm$&1.6&42.1&$\pm$&2.2&$\pm$&1.6&$\pm$&1.6\\
3&65.2&$\pm$&1.5&$\pm$&1.8&$\pm$&2.5&66.1&$\pm$&2.5&$\pm$&2.1&$\pm$&2.6\\
4&91.3&$\pm$&1.8&$\pm$&2.3&$\pm$&3.6&87.9&$\pm$&2.9&$\pm$&2.6&$\pm$&3.4\\
5&108.0&$\pm$&2.0&$\pm$&2.7&$\pm$&4.2&95.8&$\pm$&3.0&$\pm$&2.8&$\pm$&3.7\\
6&121.4&$\pm$&2.1&$\pm$&3.0&$\pm$&4.7&118.5&$\pm$&3.3&$\pm$&3.4&$\pm$&4.6\\
7&136.0&$\pm$&2.2&$\pm$&3.3&$\pm$&5.3&133.3&$\pm$&3.6&$\pm$&3.7&$\pm$&5.2\\
8&140.8&$\pm$&2.2&$\pm$&3.4&$\pm$&5.5&141.3&$\pm$&3.7&$\pm$&3.9&$\pm$&5.5\\
9&145.5&$\pm$&2.3&$\pm$&3.5&$\pm$&5.7&151.2&$\pm$&4.0&$\pm$&4.2&$\pm$&5.9\\
10&144.0&$\pm$&2.3&$\pm$&3.4&$\pm$&5.6&133.6&$\pm$&3.9&$\pm$&3.7&$\pm$&5.2\\
11&137.1&$\pm$&2.2&$\pm$&3.3&$\pm$&5.3&129.6&$\pm$&4.1&$\pm$&3.7&$\pm$&5.1\\
12&121.8&$\pm$&2.1&$\pm$&3.0&$\pm$&4.8&116.5&$\pm$&4.0&$\pm$&3.4&$\pm$&4.5\\
13&100.4&$\pm$&1.9&$\pm$&2.4&$\pm$&3.9&93.5&$\pm$&3.8&$\pm$&2.9&$\pm$&3.6\\
14&75.2&$\pm$&1.7&$\pm$&1.8&$\pm$&2.9&63.8&$\pm$&3.7&$\pm$&2.2&$\pm$&2.5\\
15&57.9&$\pm$&1.5&$\pm$&1.5&$\pm$&2.3&58.6&$\pm$&3.7&$\pm$&2.4&$\pm$&2.3\\
16&41.1&$\pm$&1.2&$\pm$&1.3&$\pm$&1.6&34.7&$\pm$&4.0&$\pm$&1.9&$\pm$&1.4\\
17&18.4&$\pm$&0.6&$\pm$&0.6&$\pm$&0.7&18.8&$\pm$&3.2&$\pm$&1.6&$\pm$&0.7\\
18&2.6&$\pm$&0.2&$\pm$&0.3&$\pm$&0.1&
\end{tabular}
\end{center}
\end{sidewaystable}

\begin{sidewaystable}
\begin{center}
\caption{The measured differential cross-sections as a function of \phist. The first uncertainty is due to the size of the dataset, the second is due to experimental systematic uncertainties, and the third is due to the luminosity.}
\label{tab:phist}
\begin{tabular}{c|ccccccc| ccccccc}
Bin index & \multicolumn{7}{c|}{$\rm{d}\csz^{\Pmu\Pmu}/\rm{d}\phist$  [pb]} & \multicolumn{7}{c}{$\rm{d}\csz^{\rm{ee}}/\rm{d}\phist$  [pb]}\\\hline
1&1873&$\pm$&29&$\pm$&45&$\pm$&73&1725&$\pm$&49&$\pm$&48&$\pm$&67\\
2&1741&$\pm$&28&$\pm$&42&$\pm$&68&1696&$\pm$&49&$\pm$&48&$\pm$&66\\
3&1635&$\pm$&27&$\pm$&39&$\pm$&64&1549&$\pm$&47&$\pm$&44&$\pm$&60\\
4&1330&$\pm$&17&$\pm$&32&$\pm$&52&1296&$\pm$&30&$\pm$&35&$\pm$&51\\
5&983&$\pm$&15&$\pm$&24&$\pm$&38&955&$\pm$&26&$\pm$&27&$\pm$&37\\
6&722&$\pm$&10&$\pm$&17&$\pm$&28&730&$\pm$&19&$\pm$&20&$\pm$&28\\
7&471&$\pm$&7&$\pm$&11&$\pm$&18&432&$\pm$&11&$\pm$&12&$\pm$&17\\
8&300&$\pm$&5&$\pm$&7&$\pm$&12&300&$\pm$&10&$\pm$&9&$\pm$&12\\
9&160.4&$\pm$&2.7&$\pm$&3.8&$\pm$&6.3&152.4&$\pm$&4.7&$\pm$&4.4&$\pm$&5.9\\
10&81.2&$\pm$&1.9&$\pm$&1.9&$\pm$&3.2&82.6&$\pm$&3.6&$\pm$&2.7&$\pm$&3.2\\
11&38.0&$\pm$&0.9&$\pm$&0.9&$\pm$&1.5&34.0&$\pm$&1.7&$\pm$&1.1&$\pm$&1.3\\
12&14.72&$\pm$&0.58&$\pm$&0.36&$\pm$&0.57&14.71&$\pm$&1.01&$\pm$&0.63&$\pm$&0.57\\
13&6.21&$\pm$&0.27&$\pm$&0.16&$\pm$&0.24&4.94&$\pm$&0.43&$\pm$&0.23&$\pm$&0.19\\
14&1.289&$\pm$&0.086&$\pm$&0.043&$\pm$&0.050&1.213&$\pm$&0.148&$\pm$&0.080&$\pm$&0.047\\
15&0.190&$\pm$&0.021&$\pm$&0.009&$\pm$&0.007&0.201&$\pm$&0.042&$\pm$&0.021&$\pm$&0.008

\end{tabular}
\end{center}
\end{sidewaystable}
\begin{table}[!t]
\begin{center}
\caption{The measured differential cross-sections as a function of \pt. The first uncertainty is due to the size of the dataset, the second is due to experimental systematic uncertainties, and the third is due to the luminosity.}
\label{tab:pt}
\begin{tabular}{c|ccccccc}
Bin index & \multicolumn{7}{c}{$\rm{d}\csz^{\Pmu\Pmu}/\rm{d}{p_\text{T, Z}}$  [pb / GeV]}\\\hline
1&5.55&$\pm$&0.11&$\pm$&0.15&$\pm$&0.22\\
2&11.01&$\pm$&0.21&$\pm$&0.29&$\pm$&0.43\\
3&11.36&$\pm$&0.21&$\pm$&0.30&$\pm$&0.44\\
4&11.06&$\pm$&0.21&$\pm$&0.29&$\pm$&0.43\\
5&9.93&$\pm$&0.18&$\pm$&0.26&$\pm$&0.39\\
6&8.86&$\pm$&0.16&$\pm$&0.23&$\pm$&0.35\\
7&7.22&$\pm$&0.13&$\pm$&0.19&$\pm$&0.28\\
8&6.48&$\pm$&0.11&$\pm$&0.18&$\pm$&0.25\\
9&5.28&$\pm$&0.09&$\pm$&0.14&$\pm$&0.21\\
10&4.29&$\pm$&0.07&$\pm$&0.12&$\pm$&0.17\\
11&2.88&$\pm$&0.05&$\pm$&0.08&$\pm$&0.11\\
12&1.760&$\pm$&0.029&$\pm$&0.046&$\pm$&0.069\\
13&0.709&$\pm$&0.011&$\pm$&0.018&$\pm$&0.028\\
14&0.0376&$\pm$&0.0009&$\pm$&0.0010&$\pm$&0.0015
\end{tabular}
\end{center}
\end{table}

\begin{sidewaystable}
   \caption{The correlation matrix for the differential cross-section measurement as a function of \PZ boson rapidity, for the dimuon final state, excluding the luminosity uncertainty, which is fully correlated between bins.} 
 
    \begin{tabular}{c | c c c c c c c c c c c c c c c c c c}
      Bin index & 1 & 2 & 3 & 4 & 5 & 6 & 7 & 8 & 9 & 10 & 11 & 12 & 13 & 14 & 15 & 16 & 17 & 18 \\ \hline
1 & 1.00& &  &  &  &  &  &  &  &  &  &  &  &  &  &  &  &  \\
2 & 0.37&1.00& &  &  &  &  &  &  &  &  &  &  &  &  &  &  &   \\
3 & 0.35&0.50&1.00& &  &  &  &  &  &  &  &  &  &  &  &  &  &  \\
4 & 0.35&0.51&0.57&1.00& &  &  &  &  &  &  &  &  &  &  &  &  &   \\
5 & 0.35&0.50&0.57&0.62&1.00& &  &  &  &  &  &  &  &  &  &  &  &  \\
6 & 0.34&0.50&0.57&0.62&0.64&1.00& &  &  &  &  &  &  &  &  &  &  &  \\
7 & 0.34&0.50&0.57&0.63&0.65&0.67&1.00& &  &  &  &  &  &  &  &  &  &   \\
8 & 0.33&0.49&0.57&0.62&0.64&0.67&0.68&1.00& &  &  &  &  &  &  &  &  &  \\
9 & 0.33&0.48&0.57&0.61&0.63&0.66&0.67&0.69&1.00& &  &  &  &  &  &  &  &   \\
10 & 0.32&0.47&0.55&0.61&0.64&0.66&0.68&0.68&0.67&1.00& &  &  &  &  &  &  &  \\
11 & 0.31&0.45&0.53&0.59&0.62&0.64&0.66&0.67&0.67&0.68&1.00& &  &  &  &  &  &  \\
12 & 0.28&0.42&0.48&0.57&0.59&0.61&0.63&0.62&0.61&0.65&0.64&1.00& &  &  &  &  &   \\
13 & 0.28&0.41&0.47&0.54&0.57&0.59&0.61&0.61&0.60&0.64&0.63&0.63&1.00& &  &  &  &   \\
14 & 0.26&0.38&0.43&0.50&0.53&0.54&0.57&0.56&0.56&0.59&0.59&0.59&0.58&1.00& &  &  &    \\
15 & 0.23&0.34&0.39&0.46&0.48&0.49&0.51&0.51&0.50&0.55&0.54&0.55&0.54&0.51&1.00& &  &   \\
16 & 0.19&0.28&0.31&0.37&0.39&0.39&0.41&0.41&0.39&0.44&0.44&0.46&0.45&0.43&0.42&1.00& &   \\
17 & 0.19&0.27&0.31&0.36&0.38&0.39&0.40&0.40&0.40&0.44&0.44&0.45&0.44&0.42&0.41&0.35&1.00& \\
18 & 0.05&0.08&0.09&0.11&0.11&0.11&0.12&0.12&0.11&0.13&0.13&0.14&0.14&0.14&0.14&0.12&0.14&1.00 

      \label{tab:corr_rap}
    \end{tabular}

\end{sidewaystable}

\begin{sidewaystable}
  \caption{The correlation matrix for the differential cross-section measurements as a function of the \PZ boson rapidity, for the dielectron final state, excluding the luminosity uncertainty, which is fully correlated between bins.}
    \begin{tabular}{c | c c c c c c c c c c c c c c c c c }
   Bin index & 1 & 2 & 3 & 4 & 5 & 6 & 7 & 8 & 9 & 10 & 11 & 12 & 13 & 14 & 15 & 16 & 17  \\ \hline
    1 &	1.00		&			&			&			&			&			&			&			&			&			&			&			&			&			&			&			&			\\
2 &	0.07		&	1.00		&			&			&			&			&			&			&			&			&			&			&			&			&			&			&			\\
3 &	0.09		&	0.19		&	1.00		&			&			&			&			&			&			&			&			&			&			&			&			&			&			\\
4 &	0.09		&	0.17		&	0.22		&	1.00		&			&			&			&			&			&			&			&			&			&			&			&			&			\\
5 &	0.11		&	0.22		&	0.28		&	0.26		&	1.00		&			&			&			&			&			&			&			&			&			&			&			&			\\
6 &	0.12		&	0.24		&	0.30		&	0.28		&	0.35		&	1.00		&			&			&			&			&			&			&			&			&			&			&			\\
7 &	0.12		&	0.24		&	0.31		&	0.29		&	0.36		&	0.39		&	1.00		&			&			&			&			&			&			&			&			&			&			\\
8 &	0.12		&	0.24		&	0.31		&	0.29		&	0.37		&	0.39		&	0.40		&	1.00		&			&			&			&			&			&			&			&			&			\\
9 &	0.12		&	0.24		&	0.31		&	0.28		&	0.36		&	0.39		&	0.40		&	0.40		&	1.00		&			&			&			&			&			&			&			&			\\
10 &	0.11		&	0.23		&	0.29		&	0.27		&	0.34		&	0.37		&	0.38		&	0.38		&	0.38		&	1.00		&			&			&			&			&			&			&			\\
11 &	0.11		&	0.22		&	0.28		&	0.26		&	0.33		&	0.35		&	0.36		&	0.36		&	0.36		&	0.34		&	1.00		&			&			&			&			&			&			\\
12 &	0.10		&	0.21		&	0.26		&	0.24		&	0.31		&	0.33		&	0.34		&	0.34		&	0.34		&	0.32		&	0.31		&	1.00		&			&			&			&			&			\\
13 &	0.09		&	0.18		&	0.23		&	0.22		&	0.27		&	0.29		&	0.30		&	0.30		&	0.30		&	0.29		&	0.27		&	0.26		&	1.00		&			&			&			&			\\
14 &	0.07		&	0.14		&	0.18		&	0.17		&	0.21		&	0.23		&	0.23		&	0.24		&	0.23		&	0.22		&	0.21		&	0.20		&	0.18		&	1.00		&			&			&			\\
15 &	0.07		&	0.14		&	0.17		&	0.16		&	0.20		&	0.22		&	0.22		&	0.23		&	0.22		&	0.21		&	0.20		&	0.19		&	0.17		&	0.13		&	1.00		&			&			\\
16 &	0.04		&	0.08		&	0.11		&	0.10		&	0.13		&	0.13		&	0.14		&	0.14		&	0.14		&	0.13		&	0.13		&	0.12		&	0.10		&	0.08		&	0.08		&	1.00		&			\\
17 &	0.03		&	0.06		&	0.07		&	0.07		&	0.09		&	0.09		&	0.10		&	0.10		&	0.10		&	0.09		&	0.09		&	0.08		&	0.07		&	0.06		&	0.05		&	0.03		&	1.00		
      \label{tab:corr_rap_electron}
    \end{tabular}
\end{sidewaystable}

\begin{sidewaystable}
   \caption{The correlation matrix for the differential cross-section measurement as a function of \phist, for the dimuon final state, excluding the luminosity uncertainty, which is fully correlated between bins.} 
    \begin{tabular}{c | c c c c c c c c c c c c c c c}
      Bin index & 1    & 2 & 3 & 4 & 5 & 6 & 7 & 8 & 9 & 10 & 11 & 12 & 13 & 14 & 15 \\ \hline
1 & 1.00 &  &  &  &  &  &  &  &  &  &  &  &  &  &  \\
2 & 0.69 & 1.00 &  &  &  &  &  &  &  &  &  &  &  &  &  \\
3 & 0.68 & 0.66 & 1.00 &  &  &  &  &  &  &  &  &  &  &  &  \\
4 & 0.73 & 0.72 & 0.71 & 1.00 &  &  &  &  &  &  &  &  &  &  &  \\
5 & 0.70 & 0.69 & 0.67 & 0.73 & 1.00 &  &  &  &  &  &  &  &  &  &  \\
6 & 0.71 & 0.70 & 0.69 & 0.74 & 0.71 & 1.00 &  &  &  &  &  &  &  &  &  \\
7 & 0.70 & 0.69 & 0.70 & 0.74 & 0.70 & 0.72 & 1.00 &  &  &  &  &  &  &  &  \\
8 & 0.67 & 0.65 & 0.66 & 0.70 & 0.66 & 0.68 & 0.69 & 1.00 &  &  &  &  &  &  &  \\
9 & 0.68 & 0.67 & 0.66 & 0.71 & 0.68 & 0.69 & 0.69 & 0.65 & 1.00 &  &  &  &  &  &  \\
10 & 0.59 & 0.58 & 0.57 & 0.62 & 0.59 & 0.60 & 0.60 & 0.57 & 0.58 & 1.00 &  &  &  &  &  \\
11 & 0.56 & 0.54 & 0.56 & 0.60 & 0.55 & 0.57 & 0.59 & 0.56 & 0.55 & 0.48 & 1.00 &  &  &  &  \\
12 & 0.43 & 0.42 & 0.42 & 0.45 & 0.42 & 0.44 & 0.44 & 0.42 & 0.42 & 0.36 & 0.36 & 1.00 &  &  &  \\
13 & 0.40 & 0.40 & 0.38 & 0.41 & 0.41 & 0.41 & 0.40 & 0.38 & 0.39 & 0.34 & 0.31 & 0.24 & 1.00 &  &  \\
14 & 0.25 & 0.23 & 0.27 & 0.28 & 0.23 & 0.26 & 0.30 & 0.28 & 0.26 & 0.21 & 0.26 & 0.18 & 0.12 & 1.00 &  \\
15 & 0.16 & 0.15 & 0.17 & 0.18 & 0.15 & 0.17 & 0.18 & 0.17 & 0.16 & 0.14 & 0.15 & 0.11 & 0.09 & 0.10 & 1.00 

      \label{tab:corr_ps_mu}
    \end{tabular}

\end{sidewaystable}

\begin{sidewaystable}
   \caption{The correlation matrix for the differential cross-section measurements as a function of \phist, for the dielectron final state, excluding the luminosity uncertainty, which is fully correlated between bins.} 
    \begin{tabular}{c | c c c c c c c c c c c c c c c}
   Bin index &	 1		&	 2		&	 3		&	 4		&	 5		&	 6		&	 7		&	 8		&	 9		&	 10		&	 11		&	 12		&	 13		&	 14		&	 15							\\\hline
1 &	1.00		&			&			&			&			&			&			&			&			&			&			&			&			&			&												\\
2 &	0.36		&	1.00		&			&			&			&			&			&			&			&			&			&			&			&			&												\\
3 &	0.35		&	0.35		&	1.00		&			&			&			&			&			&			&			&			&			&			&			&												\\
4 &	0.45		&	0.45		&	0.43		&	1.00		&			&			&			&			&			&			&			&			&			&			&												\\
5 &	0.37		&	0.37		&	0.36		&	0.46		&	1.00		&			&			&			&			&			&			&			&			&			&												\\
6 &	0.39		&	0.38		&	0.37		&	0.48		&	0.39		&	1.00		&			&			&			&			&			&			&			&			&												\\
7 &	0.39		&	0.38		&	0.37		&	0.48		&	0.39		&	0.41		&	1.00		&			&			&			&			&			&			&			&												\\
8 &	0.34		&	0.34		&	0.33		&	0.42		&	0.35		&	0.36		&	0.36		&	1.00		&			&			&			&			&			&			&												\\
9 &	0.35		&	0.34		&	0.33		&	0.43		&	0.35		&	0.37		&	0.37		&	0.33		&	1.00		&			&			&			&			&			&												\\
10 &	0.27		&	0.27		&	0.26		&	0.34		&	0.28		&	0.29		&	0.29		&	0.25		&	0.26		&	1.00		&			&			&			&			&												\\
11 &	0.25		&	0.25		&	0.24		&	0.31		&	0.26		&	0.26		&	0.27		&	0.23		&	0.24		&	0.19		&	1.00		&			&			&			&												\\
12 &	0.19		&	0.19		&	0.18		&	0.23		&	0.19		&	0.20		&	0.20		&	0.18		&	0.18		&	0.14		&	0.13		&	1.00		&			&			&												\\
13 &	0.15		&	0.15		&	0.15		&	0.19		&	0.15		&	0.16		&	0.16		&	0.14		&	0.14		&	0.11		&	0.10		&	0.08		&	1.00		&			&												\\
14 &	0.11		&	0.11		&	0.11		&	0.14		&	0.11		&	0.12		&	0.12		&	0.10		&	0.11		&	0.08		&	0.08		&	0.06		&	0.05		&	1.00		&												\\
15 &	0.08		&	0.08		&	0.08		&	0.10		&	0.08		&	0.09		&	0.09		&	0.08		&	0.08		&	0.06		&	0.06		&	0.04		&	0.03		&	0.02		&	1.00									
  
      \label{tab:corr_ps_e}
    \end{tabular}
\end{sidewaystable}

\begin{sidewaystable}
   \caption{The correlation matrix for the differential cross-section measurements as a function of boson \pt, for the dimuon final state, excluding the luminosity uncertainty, which is fully correlated between bins. } 
 
    \begin{tabular}{c | c c c c c c c c c c c c c c }
    Bin index & 1    & 2 & 3 & 4 & 5 & 6 & 7 & 8 & 9 & 10 & 11 & 12 & 13 & 14  \\ \hline
1 & 1.00 &  &  &  &  &  &  &  &  &  &  &  &  &  \\
2 & 0.54 & 1.00 &  &  &  &  &  &  &  &  &  &  &  &  \\
3 & 0.54 & 0.55 & 1.00 &  &  &  &  &  &  &  &  &  &  &  \\
4 & 0.53 & 0.54 & 0.55 & 1.00 &  &  &  &  &  &  &  &  &  &  \\
5 & 0.55 & 0.55 & 0.55 & 0.56 & 1.00 &  &  &  &  &  &  &  &  &  \\
6 & 0.54 & 0.56 & 0.56 & 0.55 & 0.56 & 1.00 &  &  &  &  &  &  &  &  \\
7 & 0.53 & 0.53 & 0.54 & 0.55 & 0.56 & 0.54 & 1.00 &  &  &  &  &  &  &  \\
8 & 0.52 & 0.54 & 0.55 & 0.55 & 0.55 & 0.55 & 0.54 & 1.00 &  &  &  &  &  &  \\
9 & 0.51 & 0.52 & 0.54 & 0.55 & 0.55 & 0.53 & 0.55 & 0.55 & 1.00 &  &  &  &  &  \\
10 & 0.53 & 0.54 & 0.55 & 0.55 & 0.57 & 0.54 & 0.55 & 0.55 & 0.56 & 1.00 &  &  &  &  \\
11 & 0.53 & 0.55 & 0.57 & 0.56 & 0.56 & 0.56 & 0.55 & 0.57 & 0.57 & 0.56 & 1.00 &  &  &  \\
12 & 0.53 & 0.56 & 0.57 & 0.56 & 0.56 & 0.56 & 0.55 & 0.58 & 0.57 & 0.57 & 0.60 & 1.00 &  &  \\
13 & 0.54 & 0.55 & 0.57 & 0.58 & 0.58 & 0.56 & 0.58 & 0.58 & 0.60 & 0.59 & 0.60 & 0.61 & 1.00 &  \\
14 & 0.42 & 0.44 & 0.45 & 0.47 & 0.47 & 0.44 & 0.47 & 0.48 & 0.50 & 0.49 & 0.50 & 0.51 & 0.54 & 1.00 
  
      \label{tab:corr_pt}
    \end{tabular}
  
\end{sidewaystable}

\clearpage

\addcontentsline{toc}{section}{References}
\setboolean{inbibliography}{true}
\bibliographystyle{LHCb}
\bibliography{main,LHCb-PAPER,LHCb-CONF,LHCb-DP,LHCb-TDR}
\newpage
                                                                    
\clearpage      
\centerline{\large\bf LHCb collaboration}
\begin{flushleft}
\small
R.~Aaij$^{39}$,
B.~Adeva$^{38}$,
M.~Adinolfi$^{47}$,
Z.~Ajaltouni$^{5}$,
S.~Akar$^{6}$,
J.~Albrecht$^{10}$,
F.~Alessio$^{39}$,
M.~Alexander$^{52}$,
S.~Ali$^{42}$,
G.~Alkhazov$^{31}$
% ...

{\footnotesize \it
% ...
$ ^{5}$Clermont Universit{\'e}, Universit{\'e} Blaise Pascal, CNRS/IN2P3, LPC, Clermont-Ferrand, France\\
$ ^{6}$CPPM, Aix-Marseille Universit{\'e}, CNRS/IN2P3, Marseille, France\\
% ...
$ ^{10}$Fakult{\"a}t Physik, Technische Universit{\"a}t Dortmund, Dortmund, Germany\\
% ...
$ ^{31}$Petersburg Nuclear Physics Institute (PNPI), Gatchina, Russia\\
% ...
$ ^{38}$Universidad de Santiago de Compostela, Santiago de Compostela, Spain\\
$ ^{39}$European Organization for Nuclear Research (CERN), Geneva, Switzerland\\
% ...
$ ^{42}$Nikhef National Institute for Subatomic Physics, Amsterdam, The Netherlands\\
% ...
$ ^{47}$H.H. Wills Physics Laboratory, University of Bristol, Bristol, United Kingdom\\
% ...
$ ^{52}$School of Physics and Astronomy, University of Glasgow, Glasgow, United Kingdom\\
% ...
}
\end{flushleft}
 
\end{document}